\documentclass[a4paper,12pt]{article}
\usepackage{float}
\usepackage{tabularx}
\usepackage{amsmath}
\usepackage{amsfonts}
\usepackage{adjustbox}
\usepackage{multirow}
\usepackage{color}
\usepackage{longtable}
\usepackage{pdflscape}
\usepackage{rotating}
\usepackage{xcolor} 
\usepackage[colorlinks = true,
            linkcolor = blue,
            urlcolor  = blue,
            citecolor = blue,
            anchorcolor = blue]{hyperref}
\usepackage{hyperref}
\hypersetup{
    urlcolor= black,                
    linkcolor= blue,                
    filecolor=black,
    citecolor=blue,
    linkbordercolor=blue
}
\usepackage[margin=0.5in,footskip=0.25in]{geometry}
\title{Pauli Consistent $\alpha$--$\alpha$ Interaction from Inverse Scattering via Phase Function Wavefunctions and RGM Antisymmetrization}
\author{Anil Khachi$^{1*}$, Shikha Awasthi$^{2}$, Tarachand Verma$^3$, Ranjana Joshi$^{1*}$ \\\\
$^{1}$Chandigarh Group of Colleges Jhanjeri, Mohali, Punjab, India- 140307\\ Chandigarh Engineering College, Department of Applied Sciences\\
$^{2}$Department of Physics and Astronomical Sciences,\\ Central University of Himachal Pradesh Dharamshala, 176215, \\Himachal Pradesh, Bharat, India\\
$^{3}$Department of Computer Science, School of Computer Science \& Engineering\\ Galgotias University, Uttar Pradesh, India.}
\begin{document}
\maketitle
\abstract{The present study employs the phase function method (PFM) to construct scattering wavefunctions for the $\alpha$--$\alpha$ system, which is central to understanding the structure of $^{8}\mathrm{Be}$. The primary objective is to analyze scattering dynamics through the reconstruction of radial wavefunctions for the $\ell = 0$, 2, and 4 partial waves within the PFM framework, thereby avoiding direct numerical integration of the Schrödinger equation.

Previously optimized single-term and two-term Morse potentials are used for benchmarking, while a double Gaussian (DG) potential is independently determined using a genetic algorithm. The resulting non-antisymmetrized wavefunctions are subsequently employed as input to the resonating group method (RGM), enabling the incorporation of Pauli exclusion effects. The antisymmetrized wavefunctions obtained in this manner show good agreement with earlier results reported by Hiura \textit{et al.}

The quasi-bound state energy for the $\ell = 0$ partial wave is evaluated using the matrix method and is found to be consistent with the experimental value of $0.08,(0.09)$~MeV. The analysis further indicates the presence of two Pauli-forbidden S-wave states, consistent with Levinson’s theorem, while the positive-energy solution near $0.08~\mathrm{MeV}$ corresponds to the physical $^{8}\mathrm{Be}$ resonance. Scattering parameters extracted from the proposed interactions are in good agreement with NLO, NNLO, and empirical results.

Overall, the results establish the effectiveness of the PFM-based framework for reconstructing scattering observables and provide further support for the robustness of phenomenological $\alpha$--$\alpha$ interaction models.}\\\\\
{\textbf{keywords:} phase function method, resonating group method, matrix method, Morse potential, Gaussian potential, phase shift, wavefunction, phase shifts, Pauli blocking effects. }

\section{Introduction}
Inverse problems occupy a central position in modern theoretical physics, particularly in quantum scattering theory, where one seeks to infer interaction properties from experimentally accessible observables. As noted by Kabanikhin, systematic investigations of inverse problems began in the 1950s in disciplines such as electrodynamics, quantum scattering, geophysics, and acoustics, and later expanded rapidly with the availability of high-performance computing resources \cite{Meoto}. Today, inverse scattering techniques play an important role across a wide range of scientific domains, including nuclear physics, where they provide a powerful framework for estimating interaction potentials using experimental phase shift data, bound state energy and differential cross section.

In recent years, renewed interest in heavy-ion and cluster scattering problems has motivated the development of advanced numerical and numerical methods to address both direct and inverse formulations of the scattering problem. Notable contributions include the numerical algorithm proposed by Gusev \textit{et al.} \cite{Gusev} for inverse optical-model analysis and the numerical investigations of direct and inverse scattering for various interaction models by Puzynina and Vo Trong Thach \cite{Puzynina}. These studies underscore the continuing relevance of inverse approaches in extracting physically meaningful information from scattering observables.

Within quantum scattering theory, the wavefunction provides the most complete description of an interaction process, as it contains all information necessary to determine measurable quantities such as scattering amplitudes, phase shifts, and cross sections. Experimentally, however, wavefunctions are not directly observable; instead, one measures asymptotic quantities such as differential and total cross sections, which are encoded in the scattering amplitude. The reconstruction of wavefunctions and amplitudes from scattering data therefore constitutes a fundamental challenge in nuclear reaction theory. Conventional approaches typically rely on solving the time-independent Schr\"odinger equation for a given interaction potential and extracting scattering information from the asymptotic behavior of the resulting wavefunctions. These methods can become computationally demanding, especially in the presence of long-range Coulomb interactions.

\noindent An alternative and efficient framework is provided by the variable phase approach (VPA), also known as the phase function method (PFM), which reformulates the scattering problem in terms of first-order nonlinear differential equations for the radial phase shift and amplitude. Unlike traditional $S$-matrix or Jost-function methods \cite{Mackintosh,Jost}, PFM requires only the interaction potential as input and avoids explicit construction of the full scattering wavefunction at the outset. The method has been successfully applied to a variety of nuclear systems, including neutron-proton \cite{AnilPRC,Scripta,NPA,reso1,reso2,ERA}, proton-proton \cite{turk,Anil}, neutron-deuteron, proton-deuteron \cite{PAN,bede1,JNP,JNP2,nD}, $n-\alpha$ \cite{lala}, $\alpha$-$\alpha$ scattering \cite{21,Khachi, compu}, $\alpha$-$^3H$ and $\alpha$-$^3He$ \cite{ijp,lala2}, $p$-$^7Be$ \cite{astro}. More recently, Laha and collaborators have demonstrated the versatility of PFM in studying nucleon-nucleon, nucleon-nucleus, and nucleus-nucleus interactions using different phenomenological potentials \cite{Laha,Khirali,Sahoo,Bhoi}.

\noindent The $\alpha$-$\alpha$ system represents an especially suitable testing ground for the phase function framework \cite{Van,Wittern}. The $\alpha$ particle is characterized by zero spin and isospin, a large binding energy of 28.3~MeV, and a small root mean square radius of about 1.44~fm, allowing it to be treated as an elementary cluster. Consequently, the $^8$Be nucleus can be viewed as a weakly bound two $\alpha$ system, and $\alpha$-$\alpha$ scattering may be modeled effectively as a two-body problem governed by a local interaction potential. Since the pioneering experimental studies of Rutherford and Chadwick in 1927, a substantial body of experimental phase shift data has become available \cite{Afzal,Chien,Igo,Darriulat,Nilson}, enabling detailed theoretical investigations of the $\alpha$-$\alpha$ interaction.

\noindent Historically, a variety of phenomenological potentials have been proposed to describe $\alpha$-$\alpha$ scattering. Early studies by H\"afner \cite{Hafner} and subsequent extensions by Nilson and collaborators \cite{Nilson} demonstrated the importance of angular-momentum dependence in reproducing experimental phase shifts. Later analysis using optical-model approaches and complex Woods-Saxon interactions revealed that no single $\ell$-independent potential can account for all partial waves over a wide energy range \cite{Igo,Darriulat}. Among simpler models, Buck \textit{et al.} \cite{Buck} employed a Gaussian nuclear interaction supplemented by a Coulomb term represented through an error function,
\begin{equation}
V(r) = -V_a e^{-\alpha r^2} + z_1 z_2 \frac{e^2}{r} \, \mathrm{erf}(\beta r),
\end{equation}
\noindent and achieved good agreement with experimental phase shifts for several even partial waves. Ali and Bodmer \cite{Ali} proposed a two-term Gaussian interaction capable of reproducing low-energy phase shifts for selected $\ell$ values, while more elaborate Woods-Saxon forms incorporating inelastic effects were later introduced for higher energies \cite{Darriulat}.

\noindent While most previous investigations have focused primarily on reproducing scattering phase shifts or cross sections, comparatively less attention has been devoted to the explicit construction and analysis of radial scattering amplitudes and wavefunctions. The phase function framework offers a natural means to address this gap, as it provides simultaneous access to the radial phase shift, amplitude, and wavefunction within a unified formalism. Motivated by the success of molecular-type interactions in describing effective forces between composite systems, the present work employs the Morse potential as a model for the short-range nuclear interaction between two $\alpha$ clusters, with the long-range Coulomb repulsion incorporated via an error-function representation.

\noindent The principal objective of this paper is to obtain a unified description of $\alpha$-$\alpha$ scattering in terms of nonantisymmetrized and antisymmetrized wavefunctions within the phase function framework. Both phase function method and resonating group method have been employed to address the absence of Pauli blocking. By employing an inverse approach constrained by high-precision experimental phase shift data, the radial evolution of the phase shift, amplitude, and wavefunction is determined for relevant partial waves. This study aims to demonstrate that PFM \cite{Calogero, Babikov}, combined with physically motivated interaction models, provides deeper insight into the dynamics of the $\alpha$-$\alpha$ system beyond conventional phase-shift analysis. The methodology adopted in the present work can be divided into six main parts:
\begin{enumerate}
\item \textit{Inverse construction of interaction potentials:} 
The interaction potentials are obtained by reproducing the experimental phase shifts using a genetic algorithm. The following interactions are considered:
\begin{enumerate}
\item Single-term Morse potential with an $erf()$ Coulomb term \cite{Khachi},
\item Two-component Morse potential \cite{PRCSastri},
\item Double Gaussian potential with screened Hulth\'en Coulomb interaction (\textit{this work}).
\end{enumerate}

\item \textit{Calculation of scattering functions:} 
The phase function $\delta_\ell(r)$ and amplitude function $A(r)$ are calculated by solving the corresponding differential equations within the framework of the PFM under appropriate boundary conditions.

\item \textit{Construction of the non-antisymmetrized wavefunction:} 
The calculated phase and amplitude functions are used to construct the non-antisymmetrized wavefunction
$
\{\chi_{\ell}^{\mathrm{PFM}}(r)\}_{\mathrm{NA}}.
$

\item \textit{Antisymmetrization within the RGM basis:} 
The non-antisymmetrized wavefunction is projected onto the antisymmetrized RGM basis through a harmonic oscillator expansion,
$
a_n = \langle \chi_{\ell}^{\mathrm{PFM}}(r)_{\mathrm{NA}} \mid R_{n\ell}(r) \rangle,
$
followed by the reconstruction
$
\Psi_\ell(r)=\sum_n a_n \mu_{n\ell} R_{n\ell}(r),
$
where $\mu_{n\ell}$ are the eigenvalues of the norm kernel accounting for Pauli blocking effects. The resulting antisymmetrized S-, D-, and G-wave functions are compared with available GCM and RGM results.

\item \textit{Low-energy scattering parameters:} 
To further examine the low-energy behavior, the scattering length and effective range are extracted for the Morse \cite{Khachi}, RPA \cite{PRCSastri}, and double Gaussian interactions. The obtained results show that the proposed interactions consistently reproduce both the scattering phase shifts and the associated low-energy observables.

\item \textit{Bound-state analysis:} 
The optimized interaction parameters are further employed in Marsiglio’s matrix method \cite{marsiglo} to determine the bound-state energy in accordance with Levinson’s theorem with a single step of $\pi$ radians \cite{levi}. Also we obtained the non-antisymmetrized wavefunction using matrix method and compared with those obtained using phase function method. 

\end{enumerate}

In summary, the non-antisymmetrized PFM wavefunction is projected onto the RGM basis to construct a fully antisymmetrized wavefunction using genetically optimized inverse interaction potentials. 
\section{Methodology: Interaction Potentials \& Phase Function Method}
\label{Sec2}
\subsubsection{Morse + \textit{erf()} Coulomb Interaction}
\label{Morse Interaction Potential}

The short-range nuclear interaction between two $\alpha$ particles is represented using the Morse potential \cite{Morse}, given by

\begin{equation}
V_{\text{Morse}}(r)=V_0
\left[
e^{-2(r-r_m)/a_m}
-2e^{-(r-r_m)/a_m}
\right],
\label{eq:morse}
\end{equation}

where $V_0$ is the potential depth, $r_m$ denotes the equilibrium separation, and $a_m$ controls the range and diffuseness of the interaction.

The Morse interaction provides a realistic phenomenological description of the effective nuclear force by incorporating short-range repulsion, intermediate-range attraction, and an exponentially decaying asymptotic behavior. Owing to these features, it has been successfully applied in several nuclear scattering studies \cite{Khirali,Bhoi,Malfiet}. In the present work, the Morse form previously employed in our earlier investigations is adopted as the nuclear interaction.

To account for the Coulomb repulsion between the two positively charged $\alpha$ particles, a finite-size error-function-modified Coulomb interaction \cite{Ali} is used:

\begin{equation}
V_C(r)=\frac{4e^2}{r}\,\mathrm{erf}(\beta r),
\label{eq:coulomb}
\end{equation}

where

\begin{equation}
\mathrm{erf}(\beta r)=
\frac{2}{\sqrt{\pi}}
\int_{0}^{\beta r} e^{-x^2}dx.
\end{equation}

The parameter $\beta$ is related to the root-mean-square radius of the $\alpha$ particle through

\begin{equation}
\beta=\frac{\sqrt{3}}{2R_\alpha}.
\end{equation}
\subsubsection{Two-Component Morse Reference Potential}

For comparison, we also employ the two-component Morse reference potential proposed by Sastri \textit{et al.} within the framework of the Reference Potential Approach (RPA) \cite{Selg1,Selg2,PRCSastri}. In this approach, the interaction is represented by two smoothly connected Morse-type functions describing the inner and outer regions of the interaction.

The potential is written as

\begin{equation}
V_{\mathrm{RPA}}(r)=
\begin{cases}
V_{\mathrm{NC}}(r), & r \le X, \\
V_{\mathrm{CL}}(r), & r \ge X,
\end{cases}
\label{eq:RPA}
\end{equation}

where $X$ denotes the matching radius. The inner-region interaction is given by

\begin{equation}
V_{\mathrm{NC}}(r)=
V_{1}+D_{1}
\left[
e^{-2\alpha_{1}(r-r_{1})}
-2e^{-\alpha_{1}(r-r_{1})}
\right],
\label{eq:VNC}
\end{equation}

while the outer-region interaction is represented by the reversed Morse form

\begin{equation}
V_{\mathrm{CL}}(r)=
V_{2}-D_{2}
\left[
e^{-2\alpha_{2}(r-r_{2})}
-2e^{-\alpha_{2}(r-r_{2})}
\right].
\label{eq:VCL}
\end{equation}

The parameters $D_1$ and $D_2$ are determined by imposing continuity of the potential and its first derivative at the matching radius $X$ \cite{PRCSastri}. In the present work, the optimized parameter set reported in Ref.~\cite{PRCSastri} is directly adopted for the calculation of $\alpha$-$\alpha$ scattering phase shifts, amplitude function, wave function (non-antisymmerized and antisymmetrized) for the $\ell=0$, $2$, and $4$ partial waves.
\subsubsection{Double Gaussian Potential+ Hulthen as screened Coulomb}
In this work we have not only compared the results by taking single term and two term Morse potential but aslo used genetic optimization to optimize the model parameters for double gaussian potential \cite{Afzal}.
\begin{equation}
V(r)=V_r.e^{-\mu_r^2 r^2}-V_a.e^{-\mu_a^2 r^2}
\end{equation}
The Coulomb interaction has been adopted using Hulthen potential \cite{majumdar}\cite{central} given as
\begin{equation}
V_C(r)=V_0\frac{e^{-r/a}}{(1-e^{-r/a})}
\end{equation}
Here $V_0$ is the strength of interaction and \textit{a} is the screening radius. $V_0$ and \textit{a} are related as $V_0 a=2K\eta$. Here K is the momentum energy in laboratory frame and $\eta$ is the Sommerfield parameter $\eta=\alpha/\hbar v$. Here \textit{v} is the relative velocity of the reactants at large seperation and $\alpha=Z_1Z_2e^2$. So, we finally have following relation between $V_0$, $a$ \& reduced mass $\mu $.
\begin{equation}
V_0 a=\frac{Z_1.Z_2e^2\mu}{\hbar^2};~~\mu=\frac{m_\alpha}{2}=1864.38525 ~MeV/c^2
\end{equation}

\subsection{Phase Function Method (PFM)}
\noindent The scattering of a spinless particle with energy $E$ and orbital angular momentum $\ell$ in a central potential $V(r)$ is governed by the radial Schr\"odinger equation
\begin{equation}
\frac{\hbar^2}{2\mu}
\left[
\frac{d^2}{dr^2}
+
\left(
k^2-\frac{\ell(\ell+1)}{r^2}
\right)
\right]
u_\ell(k,r)
=
V(r)\,u_\ell(k,r),
\label{Scheq}
\end{equation}
\noindent where $u_\ell(k,r)$ denotes the reduced radial wavefunction, $\mu$ is the reduced mass of the system, and $k=\sqrt{E/(\hbar^2/2\mu)}$ is the wave number. For the $\alpha$--$\alpha$ system, the constant $\hbar^2/2\mu$ takes the value $10.44217~\text{MeV\,fm}^2$. Under nonrelativistic kinematics, the center-of-mass energy is related to the laboratory energy through $E_{\text{CM}} = \tfrac{1}{2} E_{\ell ab}$.

\noindent Instead of solving the second-order differential equation (\ref{Scheq}) for the wavefunction, the phase function method (PFM), also known as the variable phase approach, reformulates the scattering problem in terms of a first-order nonlinear differential equation for the phase shift. This transformation relies on the mathematical equivalence between linear second-order differential equations and first-order Riccati equations, as originally developed by Calogero \cite{Calogero} and Babikov \cite{Babikov}.

\noindent Within this framework, the phase shift $\delta_\ell(k,r)$ evolves with the radial coordinate according to
\begin{equation}
\delta_{\ell}'(k,r)
=
-\frac{V(r)}{k(\hbar^2/2\mu)}
\left[
\cos\!\left(\delta_\ell(k,r)\right)\hat{j}_\ell(kr)
-
\sin\!\left(\delta_\ell(k,r)\right)\hat{\eta}_\ell(kr)
\right]^2,
\label{PFMeqn}
\end{equation}
\noindent where the prime denotes differentiation with respect to $r$, and $\hat{j}_\ell(kr)$ and $\hat{\eta}_\ell(kr)$ are the Riccati--Bessel and Riccati--Neumann functions, respectively. These functions are related to the Riccati--Hankel function of the first kind through
$\hat{h}_\ell(kr) = -\hat{\eta}_\ell(kr) + i\,\hat{j}_\ell(kr)$.
Equation~(\ref{PFMeqn}) may also be expressed in integral form as
\begin{equation}
\delta_{\ell}(k,r)
=
-\frac{1}{k(\hbar^2/2\mu)}
\int_{0}^{r}
V(r')
\left[
\cos\!\left(\delta_\ell(k,r')\right)\hat{j}_\ell(kr')
-
\sin\!\left(\delta_\ell(k,r')\right)\hat{\eta}_\ell(kr')
\right]^2
dr'.
\end{equation}
\noindent The function $\delta_\ell(k,r)$ is referred to as the \emph{phase function}, and its asymptotic value as $r \to \infty$ yields the physical scattering phase shift for the $\ell$th partial wave.

\noindent A major advantage of PFM is that the phase shifts are obtained directly from the interaction potential without requiring explicit knowledge of the wavefunction. Consequently, the numerical task is reduced to solving a single first-order nonlinear differential equation rather than a second-order boundary-value problem.

\noindent For the $s$-wave ($\ell=0$), the Riccati functions simplify to $\hat{j}_0(kr)=\sin(kr)$ and $\hat{\eta}_0(kr)=-\cos(kr)$, leading to a particularly compact phase equation,
\begin{equation}
\delta_0'(k,r)
=
-\frac{V(r)}{k(\hbar^2/2\mu)}
\sin^2\!\left[kr+\delta_0(k,r)\right].
\label{eq6}
\end{equation}

\noindent For higher partial waves, the centrifugal contribution introduces additional structure through higher-order Riccati functions. The corresponding phase equations for $\ell=2$ ($d$-wave) and $\ell=4$ ($g$-wave) are obtained by substituting the appropriate angular-momentum-dependent forms of $\hat{j}_\ell(kr)$ and $\hat{\eta}_\ell(kr)$ into Eq.~(\ref{PFMeqn}). Similar expressions can be constructed for even higher partial waves using recurrence relations for the Riccati functions.

\noindent The phase equations are numerically integrated from the origin, subject to the initial condition $\delta_\ell(0)=0$, up to sufficiently large distances where the interaction potential becomes negligible. In this asymptotic region, the phase function saturates to a constant value corresponding to the physical scattering phase shift. In the present work, the integration is carried out using a Runge-Kutta scheme of order four or five, ensuring numerical stability and accuracy across the full energy range considered.

\begin{figure*}
\centering
{\includegraphics[scale=0.82,angle=0]{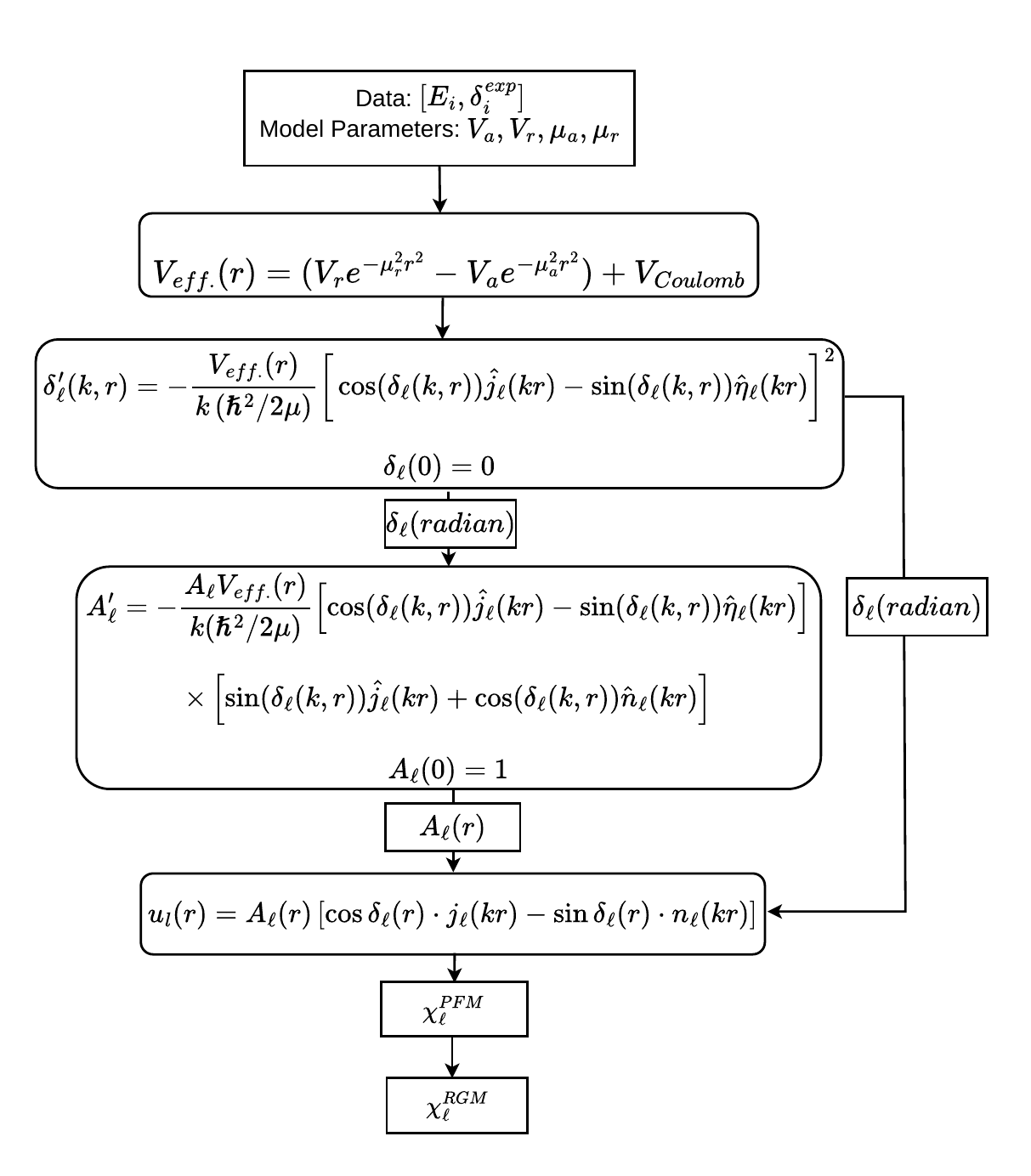}}
\caption{Detailed flowchart to obtain scattering phase shift $\delta(E)$, amplitude $A(r)$ and wavefunction $u(r)$ for $\ell=0$, 2 \& 4 waves. We took Double Gaussian (DG) plus Hulthen as potential as an effective potential. In last stage the PFM non-antisymmetrized wavefunction is fed into the RGM final antisymmetrized wavefunction.}
\label{S_r}
\end{figure*}

\subsubsection{Phase equations for $S$, $D$, and $G$ waves}
\noindent Within the phase function framework, the scattering phase shift for each partial wave evolves according to a first-order nonlinear differential equation. By substituting the appropriate Riccati-Bessel and Riccati-Neumann functions into the general phase equation, explicit expressions are obtained for different angular momentum channels.

\noindent \textbf{$s$-wave ($\ell = 0$):}\\
\noindent For the $S$-wave, the Riccati functions reduce to $\hat{j}_0(kr)=\sin(kr)$ and $\hat{\eta}_0(kr)=-\cos(kr)$. Consequently, the phase equation simplifies to
\begin{equation}
\delta_0'(k,r)
=
-\frac{V(r)}{k(\hbar^2/2\mu)}
\sin^2\!\left[kr+\delta_0(k,r)\right]
\label{eq_swave}
\end{equation}
\noindent \textbf{$d$-wave ($\ell = 2$):}\\
\noindent For the $D$-wave, the centrifugal term introduces higher-order radial contributions. The corresponding phase equation is given by
\begin{equation}
\delta_2'(k,r)
=
-\frac{V(r)}{k(\hbar^2/2\mu)}
\left[
-\sin\!\left(kr+\delta_2\right)
-\frac{3}{kr}\cos\!\left(kr+\delta_2\right)
+\frac{3}{(kr)^2}\sin\!\left(kr+\delta_2\right)
\right]^2
\label{eq_dwave}
\end{equation}
\noindent \textbf{$g$-wave ($\ell = 4$):}\\
\noindent For the $G$-wave, the phase equation involves higher-order angular momentum terms and takes the form
\begin{equation}
\begin{aligned}
\delta_4'(k,r)
&=
-\frac{V(r)}{k(\hbar^2/2\mu)}
\Bigg[
\sin\!\left(kr+\delta_4\right)
+\frac{10}{kr}\cos\!\left(kr+\delta_4\right)
-\frac{45}{(kr)^2}\sin\!\left(kr+\delta_4\right) \\
&\hspace{2.6cm}
-\frac{105}{(kr)^3}\cos\!\left(kr+\delta_4\right)
+\frac{105}{(kr)^4}\sin\!\left(kr+\delta_4\right)
\Bigg]^2 .
\end{aligned}
\label{eq_gwave}
\end{equation}
\noindent These equations are numerically integrated from the origin with the initial condition $\delta_\ell(0)=0$ up to the asymptotic region where the interaction potential vanishes. The saturation value of $\delta_\ell(k,r)$ at large $r$ yields the physical scattering phase shift for the corresponding partial wave.
\subsubsection{Amplitude Function $A_\ell(r)$}

\noindent Within the phase function framework, the scattering solution is expressed in terms of a running phase shift $\delta_\ell(k,r)$ and a radial amplitude function $A_\ell(r)$. While the phase equation governs the evolution of the local phase, the amplitude function determines the radial normalization of the scattering solution.

\noindent Starting from the phase--amplitude decomposition of the reduced radial wavefunction,

\begin{equation}
u_\ell(r)
=
A_\ell(r)
\Big[
\cos\delta_\ell(k,r)\,\hat{j}_\ell(kr)
-
\sin\delta_\ell(k,r)\,\hat{\eta}_\ell(kr)
\Big],
\label{amp_wave}
\end{equation}

\noindent where $\hat{j}_\ell(kr)$ and $\hat{\eta}_\ell(kr)$ denote the Riccati--Bessel and Riccati--Neumann functions, respectively, one obtains the following first-order differential equation for the amplitude function:

\begin{equation}
A_\ell'(r)
=
-\frac{V(r)}{k\left(\hbar^2/2\mu\right)}
A_\ell(r)
\Big[
\cos\delta_\ell(k,r)\,\hat{j}_\ell(kr)
-
\sin\delta_\ell(k,r)\,\hat{\eta}_\ell(kr)
\Big]
\Big[
\sin\delta_\ell(k,r)\,\hat{j}_\ell(kr)
+
\cos\delta_\ell(k,r)\,\hat{\eta}_\ell(kr)
\Big].
\label{amp_fun}
\end{equation}

\noindent Equation~(\ref{amp_fun}) is valid for all partial waves. The amplitude equation is integrated simultaneously with the phase equation from the origin outward subject to the initial condition

\begin{equation}
A_\ell(0)=1,
\end{equation}

\noindent which ensures regular normalization at short distances. The evolution of $A_\ell(r)$ is governed by the interaction potential $V(r)$ and modulated through the local interference between the regular and irregular free solutions. In the asymptotic region where $V(r)\rightarrow0$, the derivative $A_\ell'(r)\rightarrow0$, and the amplitude approaches a constant value. Consequently, the reconstructed wavefunction acquires the correct asymptotic scattering behavior.

\noindent For the S-wave $(\ell=0)$, the Riccati functions reduce to

\begin{equation}
\hat{j}_0(kr)=\sin(kr),
\qquad
\hat{\eta}_0(kr)=-\cos(kr).
\end{equation}

\noindent For higher partial waves $(\ell=2,4)$, the corresponding Riccati--Bessel and Riccati--Neumann functions are substituted directly into Eq.~(\ref{amp_fun}) to obtain the required amplitude functions.

\noindent The explicit expanded forms employed in the numerical implementation are provided separately in Appendix~A.

\noindent The coupled integration of the phase and amplitude equations therefore provides a unified and numerically stable procedure for reconstructing scattering wavefunctions without directly integrating the second-order radial Schr\"odinger equation.
\subsubsection{Non-antisymmetrized Wavefunction Construction}
\noindent Once the phase function $\delta_\ell(k,r)$ and the amplitude function $A_\ell(r)$ are determined, the reduced radial scattering wavefunction (non-antisymmetrized) can be reconstructed directly within the phase function framework. The solution is expressed in the phase-amplitude form as

\begin{equation}
u_\ell(r)
=
A_\ell(r)
\Big[
\cos\delta_\ell(k,r)\,\hat{j}_\ell(kr)
-
\sin\delta_\ell(k,r)\,\hat{\eta}_\ell(kr)
\Big],
\label{36}
\end{equation}

\noindent where $\hat{j}_\ell(kr)$ and $\hat{\eta}_\ell(kr)$ denote the Riccati--Bessel and Riccati--Neumann functions, respectively. This representation ensures that the reconstructed wavefunction satisfies the correct asymptotic scattering boundary condition.

\noindent In the asymptotic region $r \to \infty$, where the interaction potential $V(r)$ vanishes and the phase function saturates to its physical value $\delta_\ell(k)$, the wavefunction reduces to

\begin{equation}
u_\ell(r)
\longrightarrow
A_\ell(\infty)
\sin\!\left(
kr-\frac{\ell\pi}{2}+\delta_\ell(k)
\right),
\label{37}
\end{equation}

\noindent which corresponds to the standard asymptotic form of the partial-wave scattering solution.

\noindent For the S-wave $(\ell=0)$, the Riccati functions reduce to
$\hat{j}_0(kr)=\sin(kr)$ and
$\hat{\eta}_0(kr)=-\cos(kr)$, yielding

\begin{equation}
u_0(r)
=
A_0(r)
\sin\!\big(
kr+\delta_0(k,r)
\big).
\label{38}
\end{equation}

\noindent For higher partial waves $(\ell=2,4)$, the wavefunctions are obtained directly from Eq.~(\ref{36}) using the corresponding Riccati-Bessel and Riccati-Neumann functions.

\noindent The radial behavior of $u_\ell(r)$ therefore reflects the combined evolution of the phase and amplitude functions. In the interaction region, deviations from free-particle oscillations encode the dynamical effects of the nuclear and Coulomb interactions, while at large distances the solution smoothly approaches the asymptotic free-scattering form.

\noindent The phase function framework thus provides a direct and numerically stable procedure for constructing scattering wavefunctions without directly integrating the second-order radial Schrödinger equation. The method yields simultaneous access to phase shifts, amplitudes, and radial wavefunctions within a unified formalism.
\subsection{Antisymmetrization of PFM Wavefunction}
Recently, Xu \textit{et al.} \cite{RGM} proposed an efficient procedure to eliminate the Pauli forbidden components in the Phase Function Method (PFM) wavefunction by expanding it over a harmonic oscillator (HO) basis and utilizing the eigenvalues of the Resonating Group Method (RGM) norm kernel.
\subsubsection{HO Expansion of the Non-Antisymmetrized Wavefunction}

The non-antisymmetrized PFM wavefunction is expanded as
\begin{equation}
\{\chi_{\ell}^{\mathrm{PFM}}(r)\}_{\textit{na}}=\sum_{n} a_n R_{n\ell}(b_r;r),
\end{equation}
where $R_{n\ell}(b_r;r)$ are the radial HO wavefunctions with width parameter
\begin{equation}
b_r=\sqrt{\frac{A_1+A_2}{A_1A_2}}\,b.
\end{equation}
The expansion coefficients $a_n$ are obtained through projection
\begin{equation}
a_n = \left\langle \{\chi_{\ell}^{\mathrm{PFM}}(r)\}_{\textit{na}} \mid R_{n\ell}(b_r; r) \right\rangle,
\end{equation}
which can be written in integral form as
\begin{equation}
a_n=\int_{0}^{R_{\max}} \{\chi_\ell ^{\mathrm{PFM}}(r)\}_{\textit{na}} \, R_{n\ell}(b_r;r)\, r^2 \, dr.
\end{equation}

\subsubsection{PFM Representation via Zhaba Formalism}

Instead of solving the radial Schr\"odinger equation directly,
\begin{equation}
\frac{\hbar^{2}}{2\mu}
+
\left(
k^{2}
-
\frac{\ell(\ell+1)}{r^{2}}
\right)
\,r\,\{\chi_\ell ^{\mathrm{PFM}}(r)\}_{\textit{na}}
=
V_{\ell}(r)\,r\,\{\chi_\ell ^{\mathrm{PFM}}(r)\}_{\textit{na}},
\end{equation}
we employ the Zhaba procedure, in which the wavefunction is expressed as
\begin{equation}
\{\chi_\ell ^{\mathrm{PFM}}(r)\}_{\textit{na}}=\frac{u_{\ell}(r)}{r},
\end{equation}
where as seen already in Eq.~(\ref{36}) $u_{\ell}(r)$ is given by
\begin{equation}
u_{\ell}(r)=A_{\ell}(r)\left[\cos (\delta_\ell(k,r)) \hat{j}_{\ell}(k r)-\sin (\delta_\ell(k,r)) \hat{\eta}_{\ell}(k r)\right].
\end{equation}
For different partial waves $\ell=0, 2 ~\&~ 4$ the expression for wavefunctions is given in Appendix \ref{wavefunctions_all}.
\subsubsection{Harmonic Oscillator Basis Functions}
The radial HO wavefunctions are defined as
\begin{equation}
R_{n\ell}(b;r)
=
\mathcal{N}_{n\ell}
\left(\frac{r}{b}\right)^{\ell}
L_{n}^{\,\ell+\frac{1}{2}}
\left(\frac{r^{2}}{b^{2}}\right)
\exp\left(-\frac{r^{2}}{2b^{2}}\right),
\end{equation}
with normalization constant
\begin{equation}
\mathcal{N}_{n\ell}
=
\left[
\frac{2\,n!}
{b^{3}\,
\Gamma\left(n+\ell+\frac{3}{2}\right)}
\right]^{1/2}.
\end{equation}
\subsubsection{Antisymmetrization via RGM Kernel}
The antisymmetrized PFM wavefunction ($\{u_{\ell}^{\mathrm{PFM}}(r)\}_{\textit{as}}$), free from Pauli forbidden components, is obtained as
\begin{equation}
\{u_{\ell}^{\mathrm{PFM}}(r)\}_{\textit{as}}=
\sum_{n} a_n \sqrt{\mu_{n\ell}} \, R_{n\ell}(b_r;r),
\end{equation}
where $\mu_{n\ell}$ are the eigenvalues of the RGM norm kernel. An equivalent formulation with improved convergence, proposed by Xu \textit{et al.}, is given by
\begin{equation}
\{u_{\ell}^{\mathrm{PFM}}(r)\}_{\textit{as}}
= \{\chi_{\ell}^{\mathrm{PFM}}(r)\}_{na}
- \sum_{n} a_n \left(1 - \sqrt{\mu_{n\ell}}\right) R_{n\ell}(b_r; r).
\end{equation}
\subsubsection{Analytical Expression for RGM Eigenvalues}
Since the direct computation of $\mu_{n\ell}$ is computationally demanding, we employ analytical expressions derived by Horiuchi \cite{RGM}. For an $\alpha + x$ system, where $x$ denotes a $0s$-shell cluster ($p$, $n$, $d$, $t$, ${}^{3}\mathrm{He}$, or $\alpha$), the eigenvalues are given by
\begin{equation}
\mu_{N}
=
\frac{1}{1+\delta_{N_x,4}}
\sum_{k=0}^{N_x}
\binom{N_x}{k}
(-1)^k
\left(
1-\frac{4+N_x}{4N_x}\,k
\right)^{N}.
\end{equation}
\section{Inverse Problem Formulation}

Let the experimental phase shifts be given at discrete momenta 
\begin{equation}
\boldsymbol{\delta}^{\mathrm{exp}} = \{ \delta_\ell^{\mathrm{exp}}(k_i) \}_{i=1}^{N}.
\end{equation}
The objective of the inverse problem is to reconstruct the interaction potential \( V(r) \). We consider a parametric representation of the potential in terms of a double gaussian plus screened Coulomb form:
\begin{equation}
V(r; \theta) = V_r.e^{-\mu_r^2 r^2}-V_a.e^{-\mu_a^2 r^2}+V_0\frac{e^{-r/a}}{(1-e^{-r/a})},
\end{equation}
where the parameter vector is \( \theta = (V_r, V_a, \mu_r, \mu_a, V_0 ) \in \mathbb{R}^3 \).

Let 
\begin{equation}
\mathcal{F}_\ell : V(r) \mapsto \delta_\ell(k)
\end{equation}
denote the forward operator that maps a potential to the corresponding scattering phase shift, as obtained from the variable phase equation. The model phase shifts are then defined by
\begin{equation}
\delta_\ell^{\mathrm{model}}(k_i; \theta) = \mathcal{F}_\ell \big( V(r;\theta) \big), \quad i = 1, \dots, N.
\end{equation}

The inverse problem consists of determining the parameter vector \( \theta \) such that
\begin{equation}
\delta_\ell^{\mathrm{model}}(k_i; \theta) \approx \delta_\ell^{\mathrm{exp}}(k_i), \quad i = 1, \dots, N.
\end{equation}

This is formulated as an optimization problem:
\begin{equation}
\theta^* = \arg \min_{\theta \in \mathbb{R}^3} J(\theta),
\end{equation}
where the objective functional is defined as
\begin{equation}
J(\theta) = \frac{1}{N} \sum_{i=1}^{N}
\left|
\delta_\ell^{\mathrm{model}}(k_i; \theta) -
\delta_\ell^{\mathrm{exp}}(k_i)
\right|.
\end{equation}

\subsection{Ill-posedness of the Inverse Scattering Problem}
The inverse scattering problem consists of reconstructing an interaction potential 
\( V(r) \) from a given set of scattering phase shifts \( \delta_\ell(E) \). 
Within the phase function formulation, this can be expressed as a nonlinear operator mapping
\[
\mathcal{F}_\ell : V(r) \;\longmapsto\; \delta_\ell(E),
\]
where \( \mathcal{F}_\ell \) is defined implicitly through a Riccati-type differential equation. 
This inverse problem is well known to be \emph{ill-posed} in the sense of Hadamard, as it violates 
the conditions of stability and uniqueness.

\textbf{Instability:}
\begin{equation}
\| \boldsymbol{\delta}^{\mathrm{exp}} - \tilde{\boldsymbol{\delta}}^{\mathrm{exp}} \| \ll 1~ \text{does not imply} \quad
| V - \tilde{V} |_{L^2} \ll 1.
\end{equation}

\textbf{Non-uniqueness:}
\begin{equation}
\mathcal{F}_{\ell}(V_1) \approx \mathcal{F}_{\ell}(V_2),
\quad \text{with } V_1 \neq V_2.
\end{equation}
i.e., both potentials reproduce the same set of phase shifts within experimental accuracy. 
Such potentials are often referred to as phase-equivalent potentials. This non-uniqueness persists even when phase shifts are known over a wide energy range, unless additional constraints-such as bound-state information are imposed. Since non-uniqueness remains inherent to the present problem, we aim to identify the optimal interaction among the three considered in this work by evaluating the effect of additional conditions, namely (i) binding energy and (ii) low-energy scattering parameters. The interaction that respects these quantities can serve as a promising candidate for further clarifying the nucleus-nucleus interaction.
\subsection{Motivation for Global Optimization}

Due to the ill-posed nature of the inverse scattering problem, characterized by instability and non-uniqueness, the objective function $J(\theta)$ may exhibit multiple local minima and a highly non-convex landscape. In such cases, traditional gradient-based optimization methods may converge to suboptimal solutions and are sensitive to initial guesses.

To address these challenges, a global optimization strategy is required. The Genetic Algorithm ($GA$) is particularly suitable because it performs a population-based search and does not rely on gradient information. This allows it to explore multiple regions of the parameter space simultaneously and reduces the likelihood of being trapped in local minima.

Furthermore, the stochastic nature of $GA$ provides robustness against noise in the experimental data, which is crucial given the instability of the inverse problem.
\section{Results and Discussion}
\subsection{Optimization Procedure: Genetic Algorithm}
Like minimizing a loss function in supervised learning we here minimize a objective function or loss function or cost function by adjusting the model parameters (\textit{here for only DG potential}). The goal will be evantually be to get best fitted experimental data. The choice of the optimization method significantly affects the efficiency and convergence of a model, making it a crucial decision during the model developnment process. 

To construct the inverse potential directly from the experimental data, we optimize the model parameters of the DG potential using a metaheuristic algorithm inspired by natural selection and evolution. The ML based genetic algorithm iteratively refines the model parameters to minimize the error by reducing the loss function. Genetic algo's begin with a population of feasible solutions and evantually evolves to new set of solution at each iteration. Following steps are being employed to optimize the model parameters:
\begin{itemize}
\item \textit{Initialization:} Arrays of projectile energies $E_{\ell ab}$ and the corresponding phase shifts $\delta(E_{\ell ab})$ are given as input for analysis. The variation of SPS as a function of energy \textit{E} provides insight into emergent resonances through the analysis of the slope’s be-
havior (shown in Fig. \ref{sps_pot}). Ranges are defined for the model parameters of the DG potential to narrow the optimization space and focus on a specific region (untill cost function minimizes). To improve convergence, previously obtained parameter sets can be included in the initial population, using past solutions to search for the global optimum more efficiently.
\item \textit{Genetic algorithm:} The process starts with the creation of a random population of
parameters that represent possible solutions of the optimization routine. Each generation
involves selecting a candidate solution based on its fitness. The fitness is determined based
on the cost function, i.e. mean absolute percentage error (MAPE). The MAPE between the
simulated and experimental SPS is computed as
\[
\mathrm{MAPE} = \frac{1}{N} \sum_{i=1}^{N} \left| \frac{\delta_i^{\mathrm{exp.}} - \delta_i^{\mathrm{sim.}}}{\delta_i^{\mathrm{exp.}}} \right| \times 100
\]
where $\delta_i^{exp.}$ and $\delta_i^{sim.}$ denote the expected and simulated phase shift values at the $i^{th}$ energy
point, respectively, and \textit{N} is the total number of energy points used in the fitting procedure.
Each term in the summation represents the absolute percentage error at a specific energy.
The solution with lower \textit{MAPE} values is selected for reproduction, where genetic operators
such as crossover and mutation create new offspring. This process of selection, crossover,
and mutation continues over multiple generations, refining the model parameters to mini-
mize the cost function and improve the fit to the experimental data.
 
\item \textit{Potential determination:} The Morse potential for each parameter produced by the parent class are calculated. Following this, the parameter ranges are adjusted to ensure that
the resulting inverse potential aligns with physical constraints and meets the specified con-
ditions. Within this framework, our machine-learning-based heuristic algorithm employs the phase equation that governs the scattering process to optimize and adjust the potential
parameters.
\item \textit{Numerical solution of the phase equation:} To compute the simulated SPS, denoted as
$\delta_{sim.}$, one solves the phase equation numerically using the fifth-order Runge-Kutta (RK-5)
method. This solution uses the initialized reference potential as input, providing a simu-
lated SPS that can be compared with the expected data obtained from experimental cross
sections.
\item Testing and validation: The optimization procedure continues until the MAPE does not
change anymore, indicating that the method has successfully identified the ideal model pa-
rameters and converges to an optimal solution. Once the optimization is complete, the
inverse potential is successfully constructed with the most well optimized parameters. To
ensure reproducibility, the process can be repeated using different initial seeds. Exploring
different parameter ranges is crucial to avoid missing a solution with a lower MAPE and to
ensure the accuracy of the results.
\end{itemize}

\section{Convergence of GA with the number of generations}
\begin{figure}[H]
\centering
{\includegraphics[scale=0.7,angle=0]{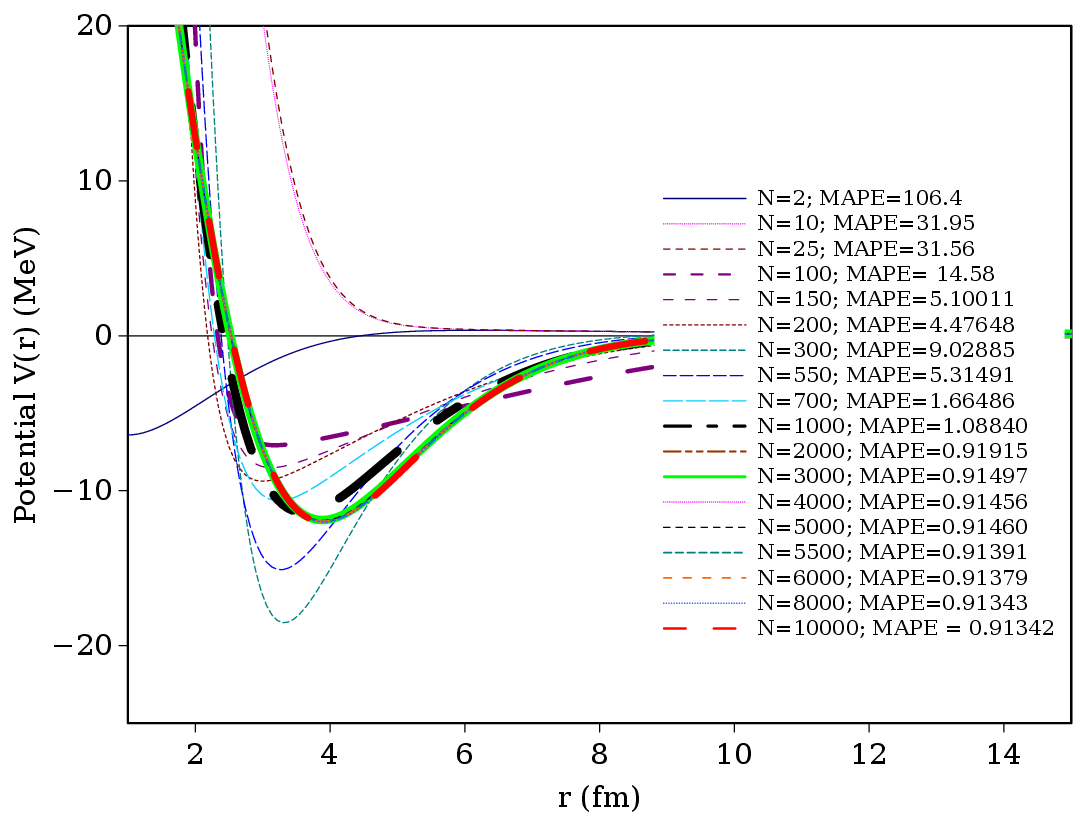}}
\caption{Visualization of potential curves obtained during optimization. As the number of generations (N)
increases, the MAPE decreases steadily, highlighting the convergence of the genetic algorithm.}
\label{Fig2}
\end{figure}

The DG potential was optimized by generating a broad parameter space with bounded parameters and sampling the initial population randomly using a fixed random seed. Figure~\ref{Fig2} displays the systematic evolution of the DG potential for $\alpha\alpha$ scattering with interparticle distance across multiple generations of the GA optimization process. Each curve represents a solution for a specific number of GA generations, with the corresponding MAPE quantifying the agreement between experimental and theoretical predictions. In the initial optimization stage ($N = 2\text{--}100$), the potential profile exhibits high MAPE values ($106\text{--}14\%$) and must be rejected, as this parameter generation fails to: (i) fit the experimental data accurately, (ii) yield low-energy scattering parameters, and (iii) reproduce the S-wave resonance energy. With a higher number of iterations ($N = 150\text{--}1000$), progressive refinement is observed, marked by lower MAPE values ($5\text{--}1\%$). Beyond $N \ge 1000$, the MAPE further decreases to $0.9\%$, significantly improving the predictive accuracy for scattering phase shifts. Convergence in MAPE occurs between $N = 5500$ and $10000$, where the potential curves overlap significantly and the MAPE stabilizes at $0.913\%$.

\section{Sensitivity analysis of model parameters}
 To overcome the limitations of the variational Monte Carlo (VMC) technique specifically its tendency to get trapped in local minima and fail to converge-we employ a genetic algorithm (GA) for global optimization. GA serves as a robust tool to navigate complex potential surfaces and reach the true global minimum. We apply this method to optimize a double Gaussian potential across varying iteration counts ($N = 2$ up to $10,000$). The mean absolute percentage error (MAPE) drops substantially from $106.4\%$ at $N=2$ to $0.9\%$ at $N=10000$, demonstrating the effectiveness of the scheme.

An evaluation of the genetic algorothm robustness and reliability in optimizing model parameters was conducted by analyzing how sensitive the results were to changes in key algorithm controls. The analysis included varying the random seed initilization, crossover probability, mutation stratergy, population size and their associated computational cost. Lets discuss these points in detail 
\begin{enumerate}
\item Random seed sensitivity and convergence characteristic: The influence of random seed
initialization on the optimization process was examined first. For relatively short evolutionary runs (fewer than $\approx$ 1000 generations), the optimized parameters exhibited noticeable dependence on the choice of seed, indicating that the algorithm tended to converge
prematurely to local minima due to insufficient exploration of the parameter space. In
contrast, when the number of generations was extended to 10000, the GA consistently
converged to the same set of optimal parameters (overlapping of N=5500-10000 potentials), independent of the initial seed. This
outcome demonstrates that, given adequate iterations, the GA reliably approaches the
global minimum in a stable and reproducible manner, rendering the final solution robust
and insensitive to stochastic initialization.
\item \textit{Role of crossover and mutation in evolutionary optimization:}  In the context of the GA,
the recombination (crossover) operator plays a critical role in combining genetic material from parent solutions to generate new offspring. Crossover probabilities (typically 0.5-0.9 i.e., 50-90\% crossover probability) promote effective mixing of information and accelerate convergence by
exploring new regions of the search space, while excessively low values reduce diversity and increase the risk of premature convergence. Conversely, mutation introduces
random alterations to individual solutions, acting as a mechanism to restore diversity and prevent the population from becoming trapped in local minima. A well calibrated balance between crossover and mutation is therefore essential: crossover exploits
promising regions of the search space, while mutation ensures exploration of alternative pathways. To maintain diversity while preserving advantageous traits, a moderate
crossover probability of 0.5 was adopted in the main optimization runs. This setting
provided a balanced trade-off between exploration and exploitation, ensuring that the
algorithm could efficiently search the parameter space and consistently approach the
global optimum. 
\item \textit{Impact of population size on model parameter convergence:} The influence of population
size on GA performance was examined through multiple runs with populations ranging
from 25 to 250. Smaller populations (≤ 100) exhibited noticeable variability, reflecting
insufficient genetic diversity and limited sampling of the search space, which often led
to premature convergence to local minima. In contrast, larger populations progressively
improved stability and reproducibility, with sizes ≥ 250 yielding nearly identical parameter values across runs. From a theoretical perspective, larger populations improve the
sampling resolution of the fitness landscape, allowing the GA to maintain a diverse set
of candidate solutions across generations. This reduces the risk of genetic drift and stagnation, thereby enhancing the probability of locating the global minimum. However,
excessively large populations can lead to diminishing returns, since the improvement in
solution quality may not scale linearly with the increased computational cost. A population size of 250 was therefore adopted for the final calculations, as it provides sufficient
diversity to ensure thorough exploration of the parameter space while remaining computationally feasible. Note: more exploration
$\rightarrow$ better convergence $\rightarrow$
more computation time.
\end{enumerate}
\renewcommand{\arraystretch}{0.9}
\begin{table}[]
\centering
\caption{GA optimized model parameters of the double Gaussian (DG) nuclear potential combined with the atomic Hulth\'en as Coulomb interaction for the $\alpha$--$\alpha$ scattering system in the partial-wave channels $l = 0$, $2$, and $4$. The parameter $a$ denotes the screening parameter associated with the Hulth\'en Coulomb potential.}

\begin{tabular}{ccccccc}
\hline 
States/Parameters & \begin{tabular}[c]{@{}c@{}}$V_a$\\ \textit{(MeV)}\end{tabular} & \begin{tabular}[c]{@{}c@{}}$\mu_a$\\$(fm^{-2})$\end{tabular} & \begin{tabular}[c]{@{}c@{}}$V_r$\\ \textit{(MeV)}\end{tabular} & \begin{tabular}[c]{@{}c@{}}$\mu_r$\\ $(fm^{-2})$\end{tabular} & \begin{tabular}[c]{@{}c@{}}$a$\\ $(fm)$\end{tabular} & \begin{tabular}[c]{@{}c@{}}MAPE\\ (\%)\end{tabular} \\ \hline
$S^+$ & 41.9889 & 0.2373 & 97.4749 & 0.4397 & 18 & 0.9134 \\ 
$D^+$ & 1530.0913 & 0.7811 & 4993.8669 & 0.9955 & 6 & 4.1955 \\ 
$G^+$ & 127.8431 & 0.4912 & 1.1949 & 4.9433 & 0.5 & 0.5416 \\ 
\hline 
\end{tabular}
\label{table1}
\end{table}  

\begin{figure}[H]
\centering
{\includegraphics[scale=0.6,angle=0]{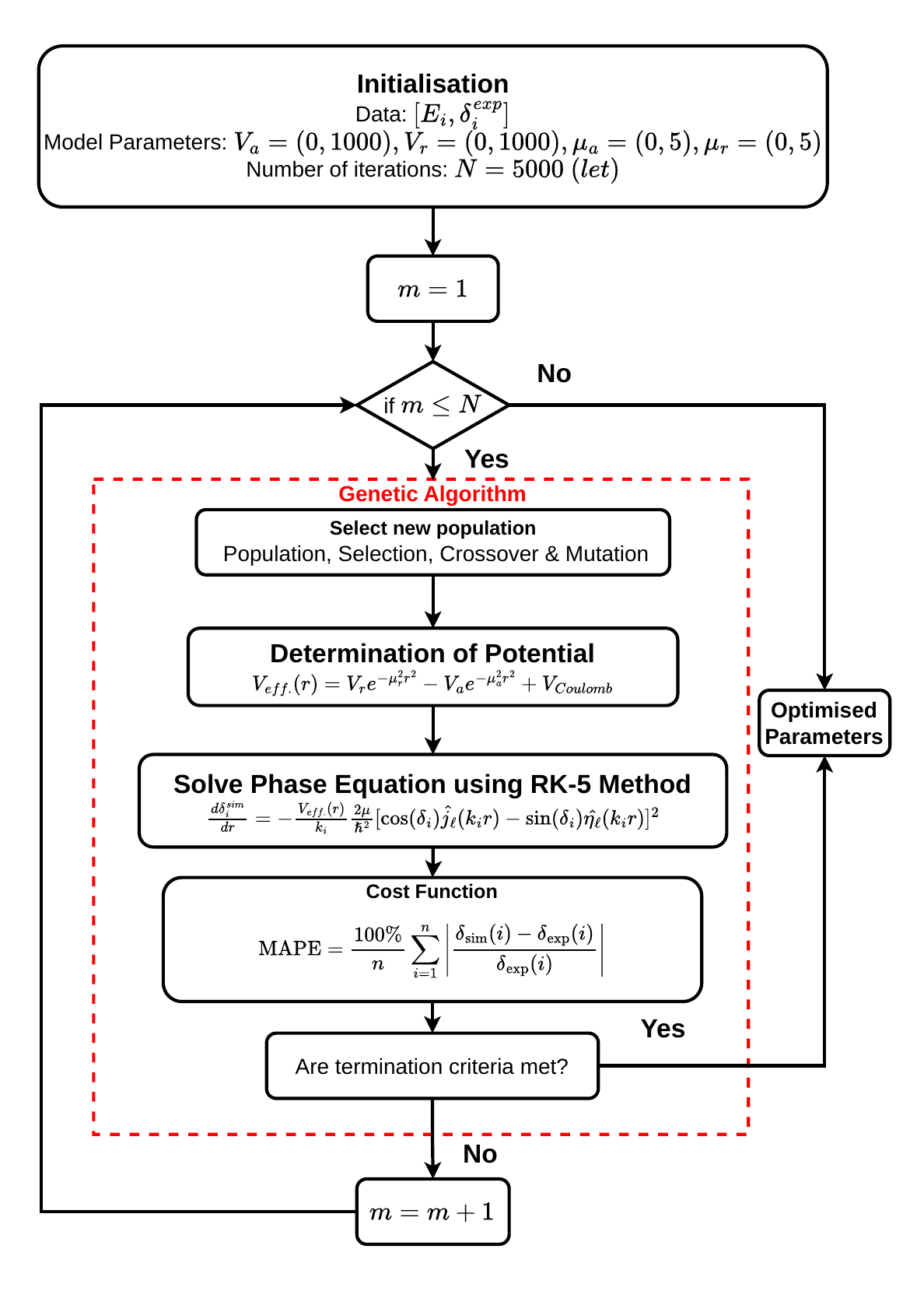}}
\caption{Flowchart illustrating the optimization procedure based on the Genetic Algorithm (GA) combined with the Phase Function Method (PFM). The procedure starts with the initialization of experimental phase-shift data and interaction model (DG) parameters, followed by the generation of new populations through selection, crossover, and mutation operations. The interaction potential is then constructed and the phase equation is solved numerically using the RK-5 method. A cost function based on the mean absolute percentage error (MAPE) is evaluated to compare the calculated and experimental phase shifts. The iterative process continues until the termination criteria are satisfied, yielding the optimized interaction parameters.}
\label{genetic}
\end{figure}
\subsection{Phase Shifts and interaction Potentials: for $\ell$=0, 2 \& 4 partial wave}
\begin{figure}[H]
\centering
{\includegraphics[scale=0.6,angle=0]{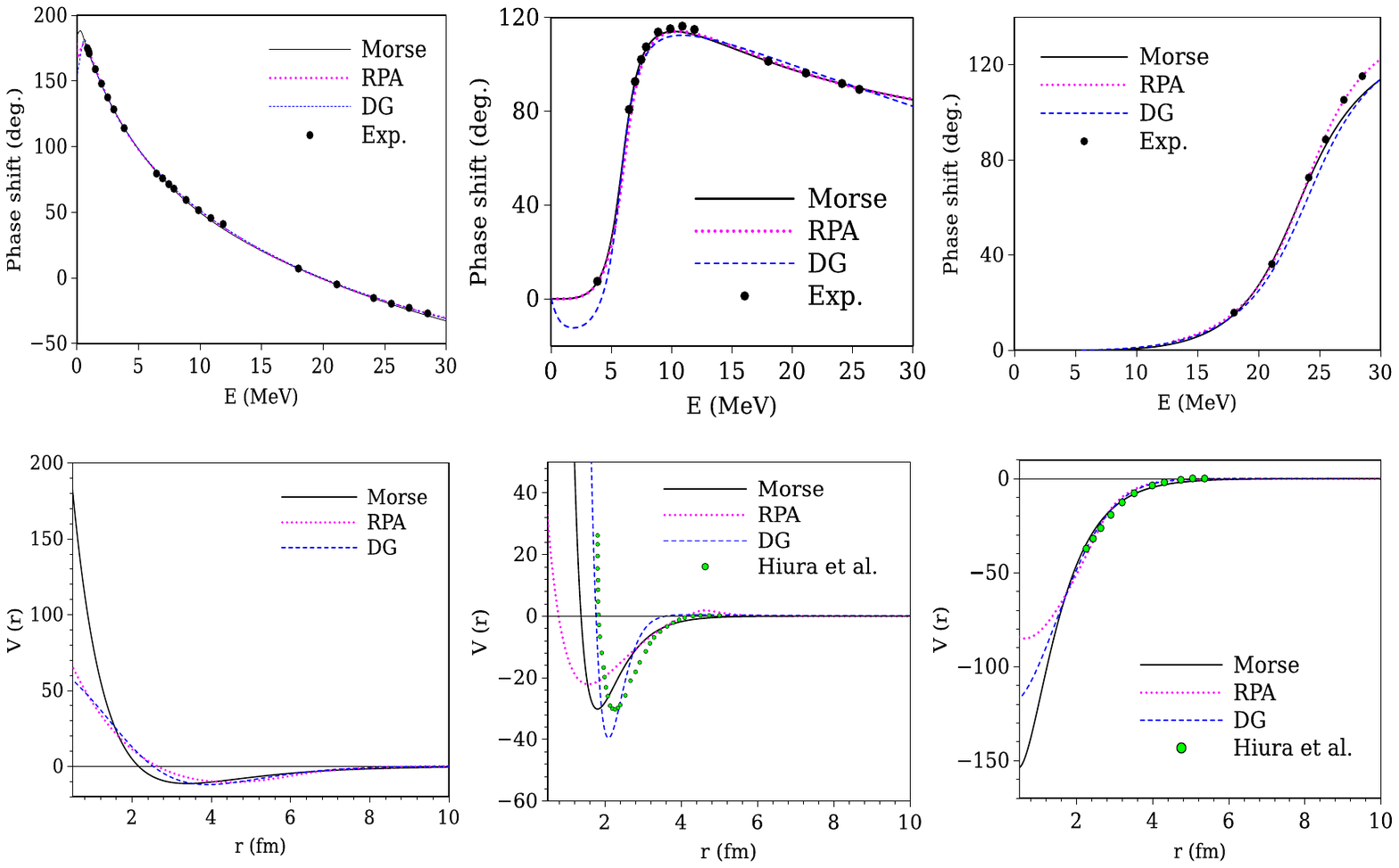}}
\caption{(upper row) Scattering phase shifts for S, D \& G cahnnels and (lower row) DG interaction potentials for the S, D \& G channels. DG interaction potential parameters are optimized using genetic algorithm. Morse \& RPA results refer to the potentials constructed using optimized parameters reported by Khachi \textit{et al.} \& Sastri et al.~\cite{Khachi}\cite{PRCSastri}. Green circles represents potentials obtained by Hiura \textit{et al.} obtained using RGM.}
\label{sps_pot}
\end{figure}

Figure \ref{sps_pot} compares the phase shifts and interaction potentials obtained from the Morse, RPA, and DG potentials for the $S$-, $D$-, and $G$-wave channels. The calculated phase shifts are shown in the top row, while the corresponding interaction potentials are presented in the bottom row.

For the $S$-wave, all three potentials reproduce the experimental phase shifts with excellent accuracy over the entire energy range. The phase shift decreases monotonically from approximately $180^\circ$ at low energies to negative values at higher energies, indicating the presence of  bound state in accordance with Levinson's theorem. Although the three interaction potentials differ significantly in their short-range repulsive cores and attractive regions, they generate nearly identical phase shifts, demonstrating the non-uniqueness of inverse scattering potentials.

The $D$-wave phase shifts also show very good agreement with the experimental data. A rapid increase is observed around $E \approx 5$--$8$ MeV, followed by a broad maximum near $10$ MeV and a gradual decrease at higher energies. The Morse, RPA, and DG potentials produce nearly indistinguishable results throughout the considered energy range. However, the corresponding interaction potentials exhibit noticeable differences in both depth and radial dependence, particularly in the attractive region around $r \approx 2$ fm. Despite these differences, the resulting scattering observables remain remarkably similar. Also note that our interactions (specially DG \& Morse) are in good aggreement with Hiura et al. interaction (green dots). Also it must be noted:
\begin{enumerate}
\item The repulsive core present in the \textit{S} and \textit{D} states, are originating from the Pauli principle,
\item The short-tailed attraction in the outside, dominated by the direct potential,
\end{enumerate}

For the $G$-wave, the phase shifts remain close to zero at low energies and increase steadily with increasing energy. The agreement between theory and experiment is satisfactory for all three potentials. The Morse and RPA results are almost identical, whereas the DG potential predicts slightly smaller phase shifts at higher energies. Although the associated interaction potentials differ substantially in depth at short distances, their influence on the scattering phase shifts is relatively weak due to the strong centrifugal barrier for $\ell=4$. Furthermore, as shown in Fig.~\ref{sps_pot}, the interaction of Hiura \textit{et al.} (green dots) aligns well with our calculations.

A comparison of the lower panels reveals that the three potentials possess significantly different radial structures, particularly at short distances where the repulsive cores and attractive wells vary considerably. Nevertheless, they reproduce nearly the same phase shifts in all three partial waves. This behavior illustrates the inverse-scattering ambiguity, where different potential forms can generate essentially identical scattering observables. The agreement obtained for the $S$-, $D$-, and $G$-wave phase shifts therefore confirms that the DG potential proposed in the present work provides a description of the $\alpha$--$\alpha$ interaction that is comparable to the previously established Morse and RPA potentials.

The present results indicate that phase-shift agreement alone is insufficient for selecting a \textit{`unique interaction potential'}. Additional observables such as scattering length, effective range, resonance energy, and binding-energy predictions must also be considered (see table \ref{scatparam}). Based on these criteria, the DG and RPA potentials provide a superior overall description compared to the Morse potential, despite all three interactions reproducing the experimental phase shifts reasonably well.

In Table~\ref{tab:common_phase_shift}, we present a comparison of scattering phase shifts obtained using the DG, Morse, and RPA interactions with experimental data. For the $\ell = 0$ channel, all three interactions exhibit excellent agreement with experiment. However, for higher partial waves, the Morse interaction yields a comparatively lower mean absolute percentage error (MAPE) than the DG and RPA interactions. Despite this improved agreement, the Morse interaction cannot be regarded as the most reliable description. As shown in Table~\ref{scatparam}, the DG and RPA interactions demonstrate better consistency for binding energy with NLO, NNLO, and empirical values, whereas the Morse interaction shows noticeable deviations.
\begin{table}[H]
\centering
\caption{Phase shifts (in degrees) for the $\alpha$--$\alpha$ system for 
$l=0$, $l=2$, and $l=4$ partial waves obtained using the DG, Morse, and RPA 
potentials in comparison with consolidated experimental data from khachi \textit{et al.} and references within \cite{Khachi}\cite{compu}\cite{Afzal}. Average $\overline{MAPE}$ is shown in the last row. }
\label{tab:common_phase_shift}
\resizebox{\textwidth}{!}{
\begin{tabular}{c|cccc|cccc|cccc}
\hline
{$E$ (MeV)}
& \multicolumn{4}{c|}{$l=0$}
& \multicolumn{4}{c|}{$l=2$}
& \multicolumn{4}{c}{$l=4$} \\
\cline{2-13}
& DG & Morse & RPA & Exp. \cite{Afzal}
& DG & Morse & RPA & Exp. \cite{Afzal}
& DG & Morse & RPA & Exp. \cite{Afzal}\\
\hline

0.85  & 175.53 & 174.28 & 175.00 & $175\pm1$
      & -10.18 & -0.01 & 0.00 & --
      & -- & -- & -- & -- \\

0.95  & 172.52 & 171.67 & 172.95 & $173\pm1$
      & -10.68 & 0.00 & 0.01 & --
      & -- & -- & -- & -- \\

1.00  & 171.13 & 170.38 & 171.84 & $171\pm1$
      & -10.89 & 0.00 & 0.01 & --
      & -- & 0.00 & -- & -- \\

1.50  & 159.16 & 158.17 & 159.75 & $159\pm1$
      & -12.16 & 0.10 & 0.13 & --
      & -- & 0.00 & -- & -- \\

2.00  & 146.94 & 147.16 & 148.00 & $148\pm1$
      & -12.38 & 0.42 & 0.45 & --
      & -- & 0.00 & -- & -- \\

2.50  & 137.12 & 137.19 & 137.38 & $137.5\pm1$
      & -11.69 & 1.12 & 1.13 & --
      & -- & 0.00 & -- & -- \\

3.00  & 127.70 & 128.08 & 127.86 & $128.4\pm1$
      & -10.07 & 2.45 & 2.39 & --
      & -- & 0.00 & -- & -- \\

3.84  & 114.20 & 114.38 & 113.91 & $114.1\pm1$
      & -4.09 & 7.21 & 6.77 & $7.50\pm1$
      & -- & 0.00 & 0.02 & -- \\

6.47  & 81.65 & 80.78 & 81.10 & $79.5\pm2$
      & 78.33 & 79.22 & 73.40 & $80.80\pm2$
      & -- & 0.07 & 0.20 & -- \\

6.96  & 76.69 & 75.69 & 76.20 & $75.9\pm3$
      & 91.28 & 92.89 & 88.37 & $92.70\pm3$
      & -- & 0.11 & 0.27 & -- \\

7.47  & 71.97 & 70.71 & 71.39 & $71.4\pm4$
      & 99.71 & 101.99 & 98.89 & $102.10\pm4$
      & -- & 0.16 & 0.36 & -- \\

7.88  & 68.35 & 66.91 & 67.72 & $68\pm4$
      & 104.08 & 106.65 & 104.42 & $107.50\pm4$
      & -- & 0.21 & 0.45 & -- \\

8.87  & 59.97 & 58.44 & 59.46 & $59.4\pm4$
      & 109.77 & 112.31 & 111.42 & $113.80\pm3$
      & -- & 0.39 & 0.73 & -- \\

9.88  & 52.34 & 50.69 & 51.80 & $51.6\pm4$
      & 111.96 & 113.97 & 113.81 & $115.20\pm2$
      & -- & 0.68 & 1.14 & -- \\

10.88 & 45.37 & 43.76 & 44.83 & $45.6\pm4$
      & 113.45 & 113.77 & 114.03 & $116.30\pm2$
      & 1.79 & 1.12 & 1.68 & -- \\

11.88 & 38.87 & 37.45 & 38.40 & $41\pm4$
      & 112.05 & 112.65 & 113.17 & $114.90\pm2$
      & 15.88 & 1.75 & 2.42 & -- \\

18.00 & 7.67 & 7.13 & 7.16 & $7.14\pm0.15$
      & 103.32 & 101.31 & 101.91 & $101.38\pm0.21$
      & 21.10 & 15.17 & 14.79 & $15.86\pm0.12$ \\

21.12 & -4.54 & -5.14 & -4.99 & $-4.96\pm0.14$
      & 97.78 & 95.99 & 96.21 & $96.33\pm0.07$
      & 72.76 & 37.07 & 33.10 & $36.27\pm0.07$ \\

24.11 & -14.62 & -15.54 & -14.92 & $-15.33\pm0.16$
      & 92.43 & 91.69 & 91.60 & $91.86\pm0.13$
      & 65.05 & 72.76 & 72.76 & $72.61\pm0.16$ \\

25.50 & -18.86 & -19.98 & -19.05 & --
      & 89.97 & 89.93 & 89.76 & $89.37\pm1.54$
      & 81.15 & 86.15 & 90.30 & $88.64\pm1.77$ \\

25.55 & -19.01 & -20.13 & -19.20 & $-19.64\pm1.01$
      & 89.88 & 89.87 & 89.70 & --
      & 81.69 & 86.25 & 98.70 & -- \\

26.99 & -23.12 & -24.46 & -23.17 & $-22.81$
      & 87.35 & 88.18 & 88.03 & --
      & 95.55 & 98.36 & 104.97 & $105.29$ \\

28.50 & -27.23 & -28.73 & -27.04 & $-27.03$
      & 84.74 & 86.51 & 86.50 & --
      & 106.37 & 107.38 & 115.36 & $115.26$ \\

\hline 
\textit{$\overline{MAPE}$} & \textit{0.21} & \textit{0.12} & \textit{0.28} & 
      & \textit{13.90} & \textit{1.13} & \textit{2.90} &
      & \textit{22.69} & \textit{3.83} & \textit{8.55
      } &  \\
\hline
\end{tabular}}
\end{table}

A key limitation of the Morse interaction arises from the use of the Coulomb term expressed via the error function, $\mathrm{erf}(r)$, which leads to an effectively long-ranged behaviour. As a result, a radial cut-off must be introduced to ensure the numerical stability and validity of the phase function method (PFM) integration. This introduces an additional model dependence that is not inherent to the physical interaction. In contrast, the DG and RPA interactions, which incorporate screened Hulthén-type potentials, exhibit naturally short-ranged behaviour. This ensures that the PFM equations remain well-defined and stable without requiring any artificial truncation, thereby preserving the internal consistency of the formalism.

Furthermore, binding energy calculations for the $\ell = 0$ state, which have not been reported in earlier works for Morse and RPA interactions, are carried out in the present study. The obtained values are $-0.138~\mathrm{MeV}$ for the Morse interaction, $0.103~\mathrm{MeV}$ for DG, and $0.073~\mathrm{MeV}$ for RPA, in comparison with the experimental value of $0.0918~\mathrm{MeV}$. The DG and RPA results are thus in closer agreement with the accepted value (see Table~\ref{scatparam}).

\begin{figure}[H]
\centering
{\includegraphics[scale=0.525,angle=0]{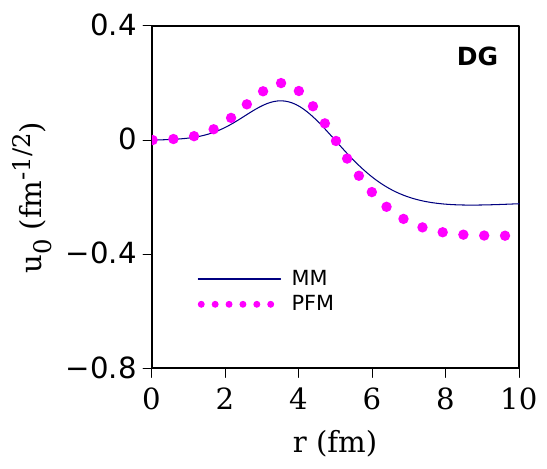}}
{\includegraphics[scale=0.52,angle=0]{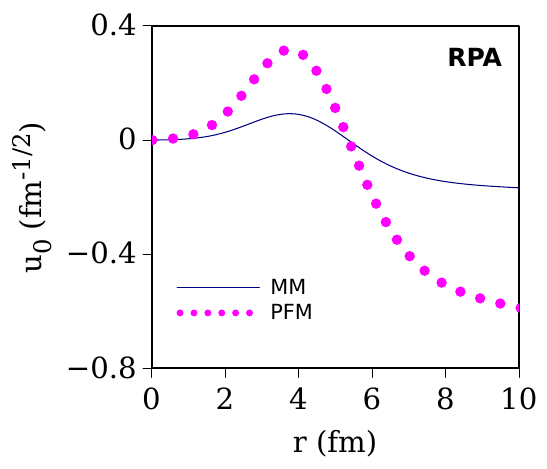}}
{\includegraphics[scale=0.5,angle=0]{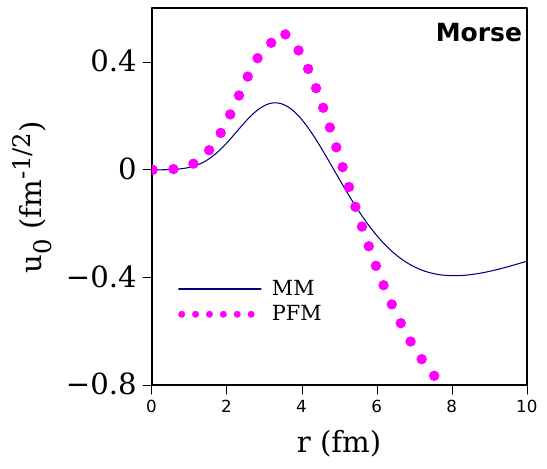}} \caption{Non-antisymmetric wavefunction for $S$-wave obtained using matrix method (MM) and PFM for three different interactions.}
\label{MM_PFM}
\end{figure}

Additional validation is provided through a comparison of wavefunctions computed using the matrix method (MM) and the PFM (see Fig. \ref{MM_PFM}). The DG interaction shows good agreement between the two methods, indicating a high degree of internal consistency, while the RPA interaction also demonstrates good agreement. In contrast, the Morse interaction exhibits noticeable deviations between MM and PFM wavefunctions, suggesting limitations in its ability to provide a unified description across different computational approaches.

Overall, while the Morse interaction may reproduce certain experimental observables with high accuracy, its reliance on a long-range Coulomb form requiring artificial truncation, along with inconsistencies in wavefunction reconstruction, limits its robustness. The DG and RPA interactions, by contrast, provide a more consistent, stable, and physically reliable framework for describing $\alpha$--$\alpha$ scattering and bound-state properties, and can therefore be regarded as more suitable candidates for such studies.

\begin{figure}
\centering
{\includegraphics[scale=0.6,angle=0]{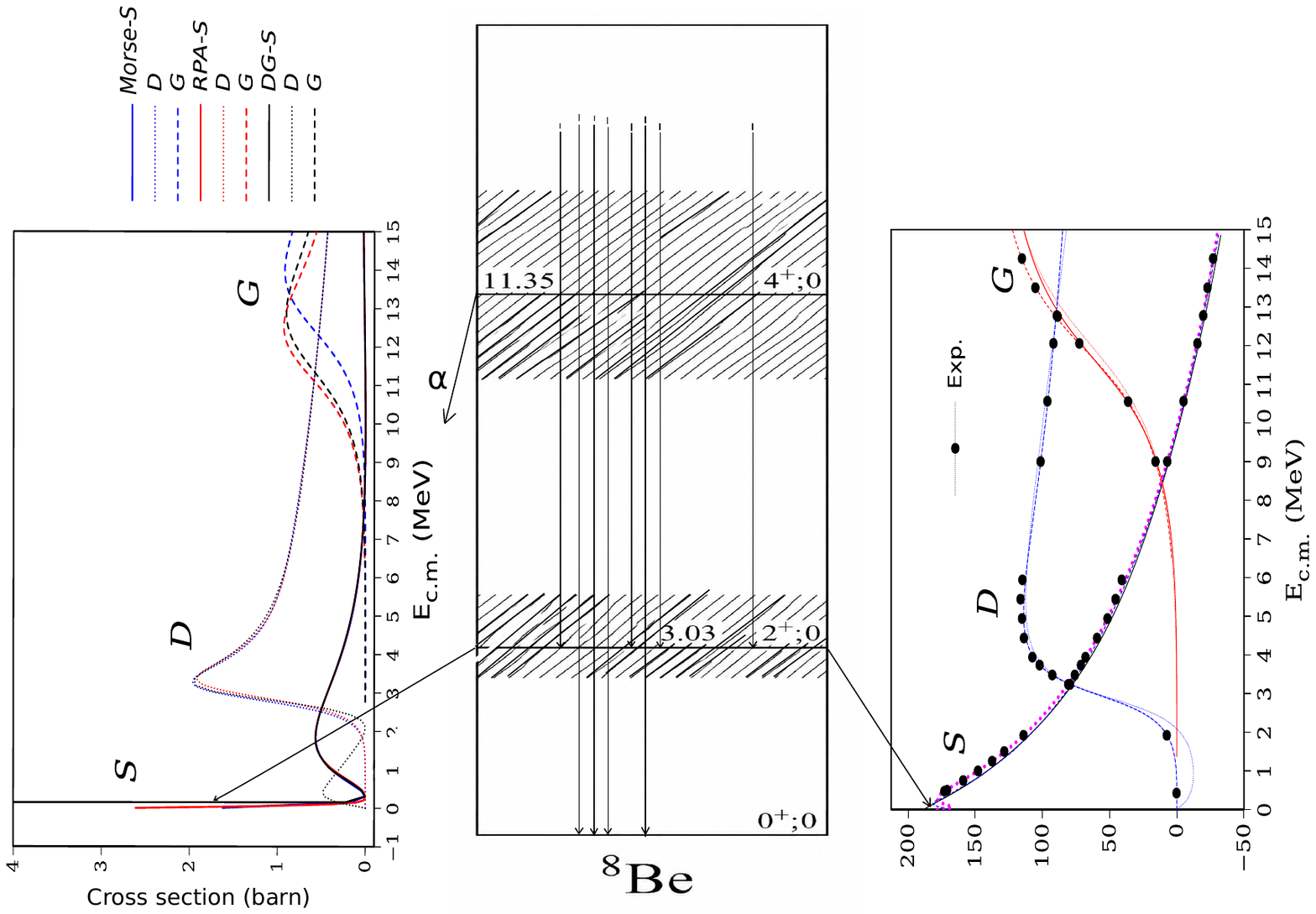}}
\caption{
Schematic representation of the relationship between the $\alpha$--$\alpha$ scattering observables and the resonance structure of the $^{8}\mathrm{Be}$ nucleus. The central panel (\textit{experimental output} \cite{link}) shows the rotational energy-level scheme of $^{8}\mathrm{Be}$ consisting of the $0^{+}$ ground state, the $2^{+}$ excited state at $3.03$ MeV, and the $4^{+}$ excited state near $11.35$ MeV. The shaded diagonal regions indicate the finite resonance widths associated with the continuum nature of these states. The left side (\textit{our work}) schematically represents the resonance contributions to the scattering cross section, while the right side (\textit{our work}) corresponds to the associated partial-wave phase shifts for the $\alpha$--$\alpha$ system. The observed resonances are connected with the dominant partial waves $l=0$, $l=2$, and $l=4$, respectively.
}
\label{fig:8Be_scheme}
\end{figure}
The energy-level scheme shown in Fig.~\ref{fig:8Be_scheme} illustrates the relationship between the experimentally observed resonant states of the $^{8}\mathrm{Be}$ nucleus and the corresponding $\alpha$--$\alpha$ scattering observables. The central panel represents the rotational resonance structure of $^{8}\mathrm{Be}$, consisting of the $0^{+}$ ground state, the $2^{+}$ excited state at $3.03$ MeV, and the $4^{+}$ excited state near $11.35$ MeV. The shaded diagonal regions indicate the finite widths of these resonances, emphasizing their continuum nature and unstable character in the $\alpha+\alpha$ channel.

The left side of the Fig.~\ref{fig:8Be_scheme} schematically represents the resonance contributions to the scattering cross section, while the right side connects these resonances with the corresponding partial-wave phase shifts obtained from the $\alpha$--$\alpha$ scattering analysis. In particular, the $0^{+}$, $2^{+}$, and $4^{+}$ states are associated with the dominant $\ell=0$, $\ell=2$, and $\ell=4$ partial waves, respectively. The resonance energies coincide with the rapid variation of the corresponding phase shifts, which is a characteristic signature of resonant scattering behavior. The figure therefore provides a unified representation of the connection between the $\alpha$--$\alpha$ phase shifts, resonance structures in the scattering cross section, and the rotational excitation spectrum of the $^{8}\mathrm{Be}$ nucleus. Such correspondence supports the cluster interpretation of $^{8}\mathrm{Be}$ as a two-$\alpha$ system and demonstrates how resonance information extracted from scattering observables can be directly related to the nuclear level structure.

\subsection{Various Scattering Functions: $\delta(r)$, A(r) \& u(r)}
Figure \ref{all} represents the variation of phase shift $\delta(r)$, amplitude function \textit{A(r)} and non-antisymmetrized wavefunction $u_0(r)$ for \textit{S, D} and \textit{G} wave respectively. The calculations are performed at their respective resonant energies. 
\begin{itemize}
\item \textit{S-wave:} The radial phase shift $\delta(r)$ exhibits a sharp increase near $r \approx 4$ fm. According to Levinson's theorem, a phase shift jump of approximately $\pi$ is associated with the presence of at least one bound (or forbidden) state. Beyond 4 fm, the phase shift approaches its asymptotic value and no further abrupt variation is observed. All three interaction potentials display a similar overall behavior, indicating a consistent description of the low-energy $S$-wave scattering process. For the DG potential, a small additional increase in the phase shift is observed beyond 8 fm. A noticeable kink appears in the amplitude function $A(r)$ just below 4 fm, which originates from the rapid variation of the scattering phase shift discussed above. The corresponding non-antisymmetrized $S$-wave functions, shown in the bottom-left panel, exhibit similar qualitative behavior for all three potentials prior to the implementation of the Pauli-blocking condition.

\item \textit{D-wave:} The phase shift remains nearly constant up to $r \approx 3$ fm, after which the three potentials begin to exhibit different radial behaviors. The RPA and DG potentials show a relatively smooth and gradual variation, whereas the Morse potential displays a more rapid decrease. This difference arises from the cutoff condition imposed on the Morse potential during the numerical integration of the Phase Function Method (PFM), which was introduced to maintain numerical stability without significantly affecting the calculated scattering observables. The amplitude functions obtained from the three potentials are found to be very similar over the entire radial range. Likewise, the non-antisymmetrized wave functions corresponding to the DG and RPA potentials are almost identical, while the Morse result shows a small deviation that can be attributed to the same cutoff prescription employed during the integration procedure.
\begin{figure}[H]
\centering
{\includegraphics[scale=0.55,angle=0]{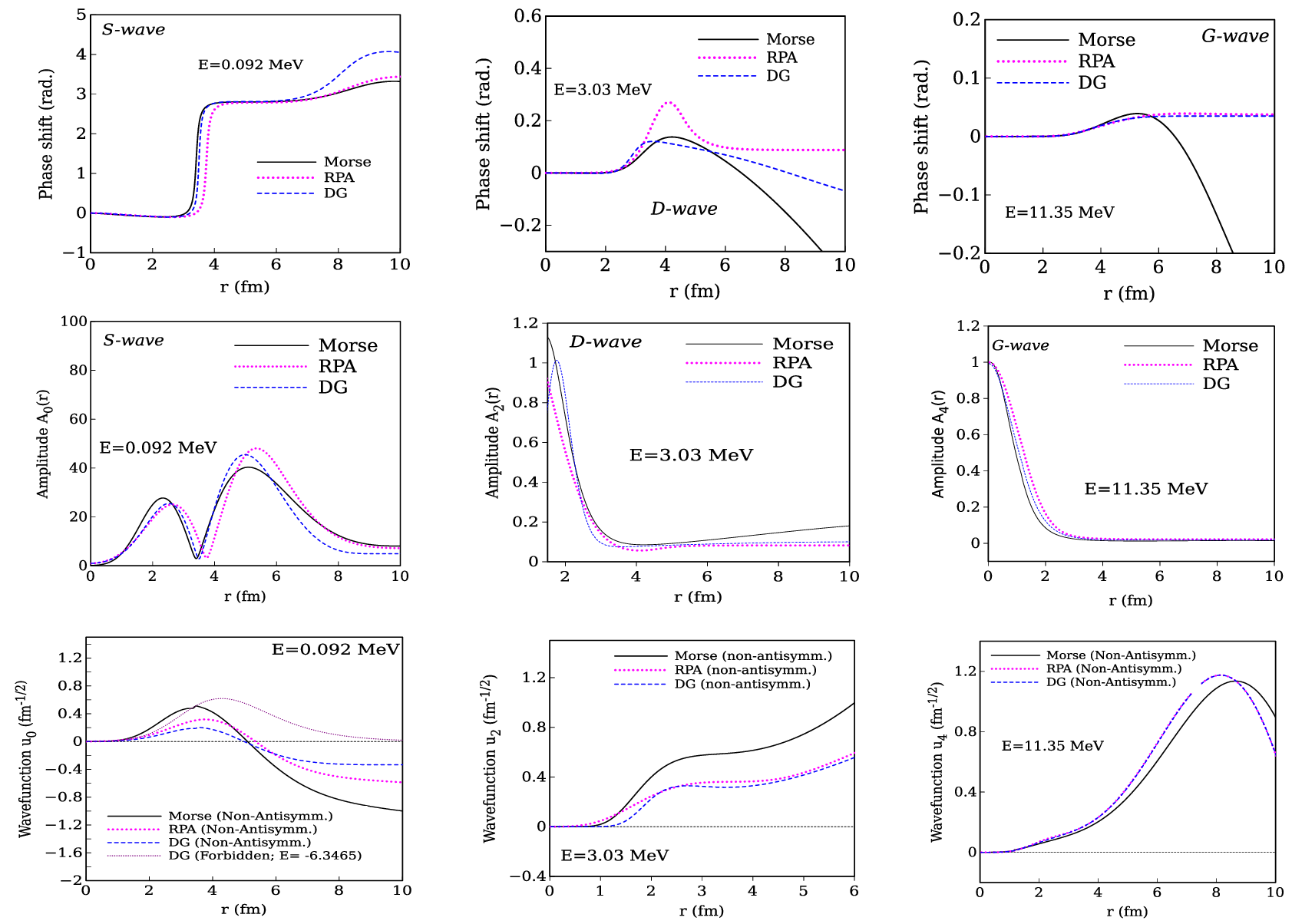}}
\caption{(top) Variation of the phase shift (in radians) as a function of distance (in fm) for the $\ell = 0$, 2, and 4 partial waves at resonant energy. RPA results correspond to the reference potential approach uiguresing the parameters reported by Sastri et al.~\cite{PRCSastri}. (mid) Amplitude functions for \textit{S, P \& D}-waves. (bottom) All the wavefunctions are calculated at resonance energies of 0.092 MeV, 3.03 MeV \& 11.35 MeV for $\ell=0, 2 ~\& ~4$ states respectively. Forbidden state non-antisymmetrized wavefunction for $\ell=0$ is shown (dotted violet) at $E=-6.3465 MeV$.}
\label{all}
\end{figure}

\item \textit{G-wave:} The $G$-wave phase shift remains nearly constant and close to zero over most of the radial range for the DG and RPA potentials, indicating a weak interaction in this partial wave. In contrast, the Morse potential exhibits a pronounced decrease beyond $r \approx 6$ fm, eventually reaching negative values. Similar to the $D$-wave case, this behavior can be attributed to the cutoff condition imposed during the PFM integration. The amplitude functions for all three potentials decay rapidly and become negligible beyond approximately 3 fm, demonstrating that the higher-angular-momentum centrifugal barrier strongly suppresses the wave amplitude at larger distances. The non-antisymmetrized $G$-wave functions obtained from the three potentials are nearly identical throughout the radial range, suggesting that the wave function is relatively insensitive to the details of the interaction potential in this partial wave.
\end{itemize}

\subsection{Levinson's Theorem}
The step-like structure observed in the phase function $\delta(r)$ at low momentum ($k = 0.01$) reflects the gradual accumulation of phase as a function of the radial coordinate. As $r \to \infty$, the phase shift approaches $\delta \approx \pi$, which, according to Levinson's theorem, indicates the presence of a single $S$-wave bound (or Pauli-forbidden) state. 

This interpretation is consistent with the matrix method (MM) diagonalization results, which yield one negative-energy eigenvalue corresponding to such a forbidden state. The next eigenvalue lies very close to the continuum threshold and is consistent with the known experimental value of $0.0918~\mathrm{MeV}$~\cite{exp}. This corresponds to a quasi-bound (resonant) state representing the physical $S$-wave resonance of the $\alpha$--$\alpha$ system, associated with the ground state of $^{8}\mathrm{Be}$. This state is not a true bound state but rather a resonance embedded in the continuum, characterized by a finite lifetime and observable through scattering.
\begin{figure}[H]
\centering
{\includegraphics[scale=0.9,angle=0]{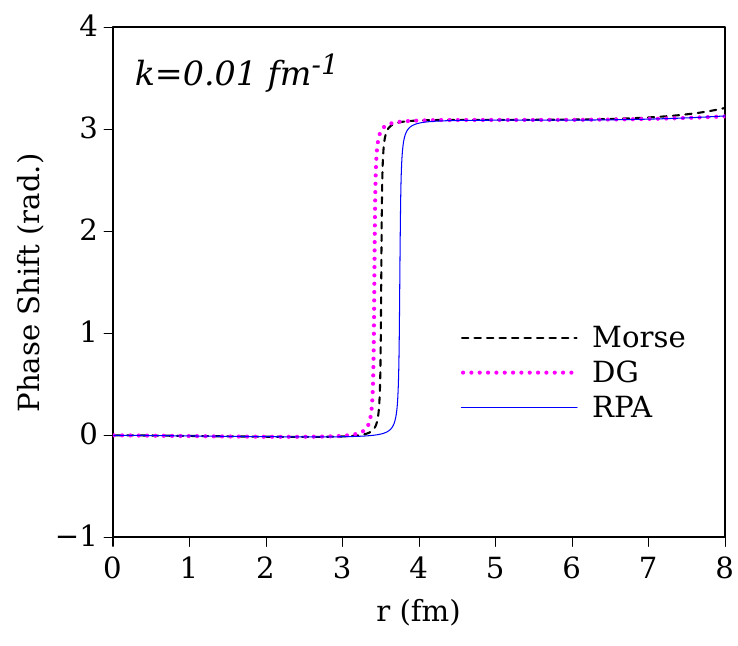}}
\caption{Phase shift variation with distance for the s-wave (Morse, DG \& RPA) at low values of $k~(fm^{-1})$. A single step of approximately $\pi$ radians is observed below $4\,\mathrm{fm}$, indicating the possibility for the presence of one bound state in accordance with Levinson's theorem.}
\label{Figxx}
\end{figure}

\subsection{Non-antisymmetric and anti-symmetric wavefunctions: Combining PFM+RGM}
Earlier in Fig. \ref{all} we extracted an effective interaction from scattering data using PFM and use it to compute relative motion wavefunctions. However, since this approach ignores the internal fermionic structure of the clusters, the resulting wavefunctions contain unphysical components. In short ``\textit{PFM provides an effective description of scattering observables, it does not incorporate Pauli blocking effects, which are naturally included in the RGM framework via antisymmetrization}''. Therefore, we will now apply RGM to antisymmetrize the wavefunctions at the nucleonic level, which enforces the Pauli principle and removes these inconsistencies. 

\begin{figure}[H]
\centering
{\includegraphics[scale=0.6,angle=0]{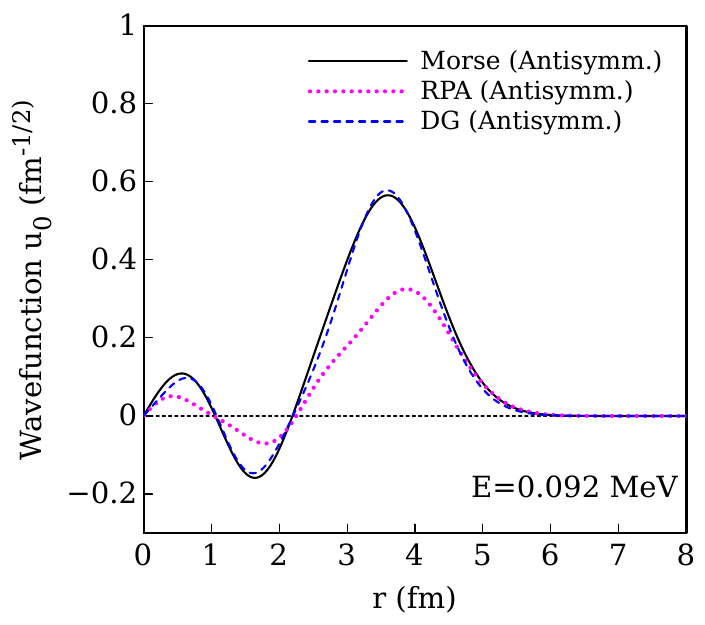}}
{\includegraphics[scale=0.62,angle=0]{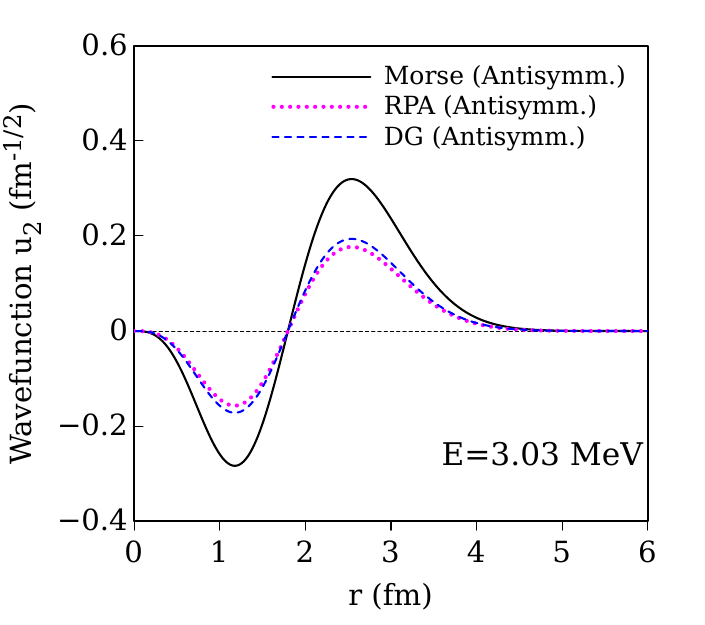}}
{\includegraphics[scale=0.6,angle=0]{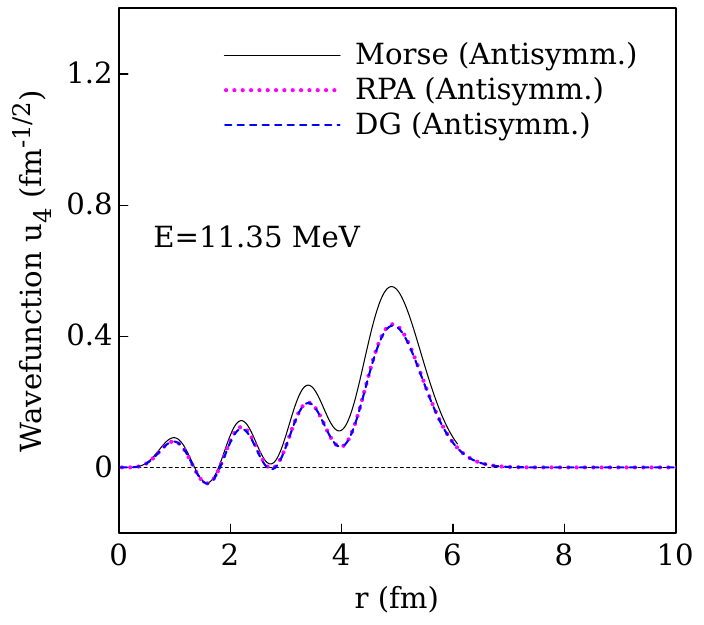}}
\caption{\textit{S,D \& G} state antisymmetrized wavefunction using Morse, reference potential approach (RPA) and double gaussian (DG) potentials at energy of E=0.092 MeV, E=3.03 MeV and E=11.35 MeV.}
\label{fig9}
\end{figure}

The antisymmetrized wavefunctions provide a direct representation of the spatial structure of the resonant $\alpha$--$\alpha$ states. The nodal behavior seen in Fig. \ref{fig9} reflects the exclusion of Pauli-forbidden configurations, while the locations of the dominant peaks indicate the most probable inter-cluster separations. The close agreement between the DG and RPA wavefunctions, particularly in the $D$- and $G$-wave channels, demonstrates that the physical resonance structure is largely insensitive to the detailed form of the interaction potential once the Pauli principle is properly incorporated.

The oscillatory structure observed in the antisymmetrized G-wave functions arises from the combined effects of the strong centrifugal barrier associated with $\ell=4$ and the orthogonality constraints imposed by the Pauli principle. The resulting wavefunctions exhibit multiple nodes and maxima, corresponding to different allowed relative-motion configurations of the two $\alpha$ clusters. The nearly identical nodal structure predicted by the Morse, RPA, and DG potentials indicates that the spatial characteristics of the G-wave resonance are largely determined by angular-momentum and antisymmetrization effects rather than by the detailed form of the interaction potential.

It is interesting to note that amplitude of the inter-cluster wave function of the \textit{S} resonance state is greater in the inner region (r$\approx$0) than in the outer space (Fig. \ref{origin}~(left)). Thus the Pauli principle allows two alpha particles to be at the same point of the coordinate space. Due to the centrifugal barrier, wave
functions of the \textit{D} and \textit{G} resonance states equal zero at r=0 fm (Fig. \ref{origin}). Other interesting feature
of all resonance states is that they have large amplitude of wave function in the internal
region ($0\leq r<7$ fm) and small amplitude of oscillations in asymptotic region ($r>7
$fm). 

\begin{figure*}
\centering
{\includegraphics[scale=0.7,angle=0]{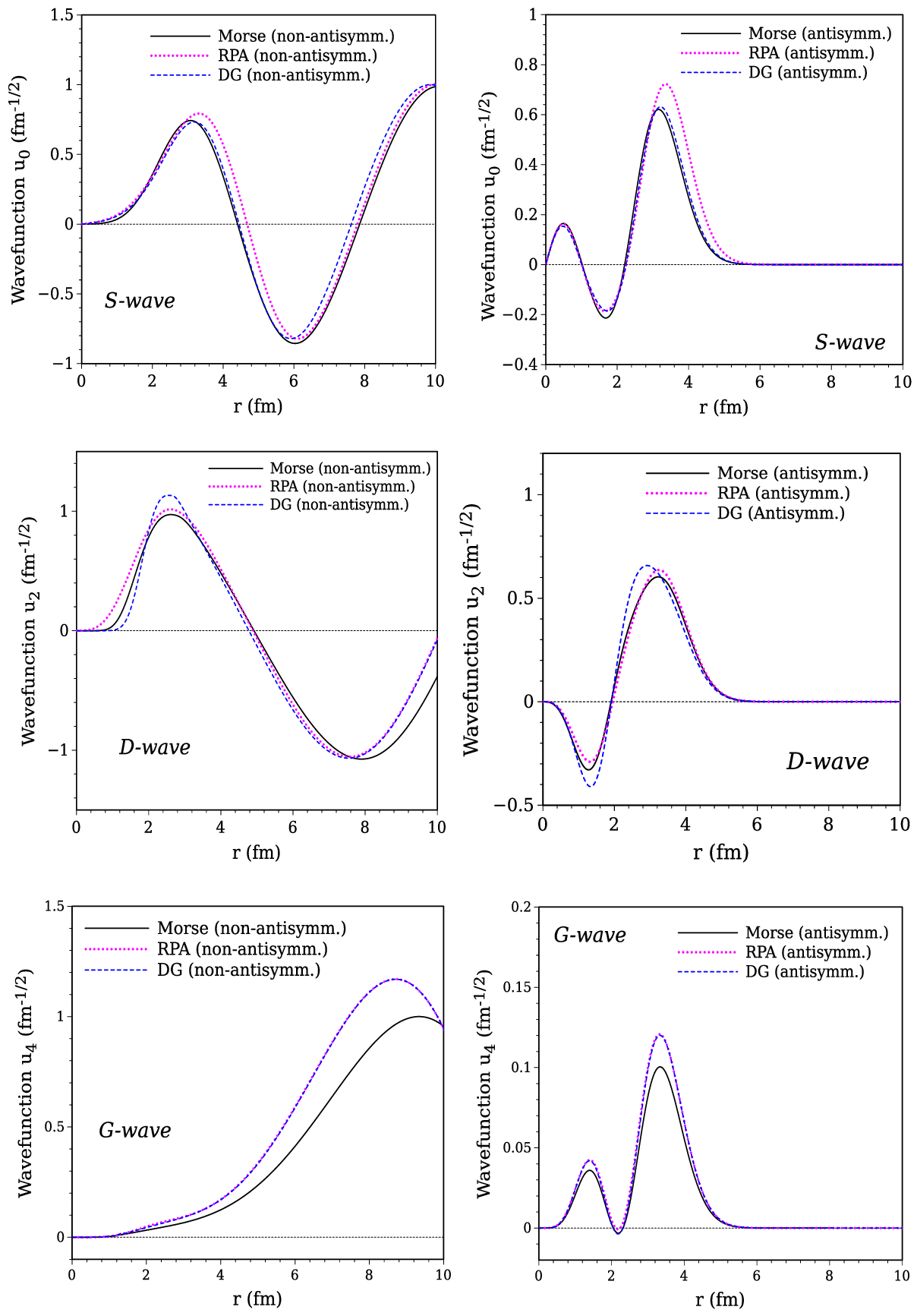}}
\caption{\textit{S, D \& G} state relative scattering wave functions (Morse, RPA and DG) before antisymmetrization (\textit{left panel}) and after antisymmetrization (right panel) calculated at E=10 MeV.}
\label{non_anti}
\end{figure*}

\begin{figure}[H]
\centering
{\includegraphics[scale=0.8,angle=0]{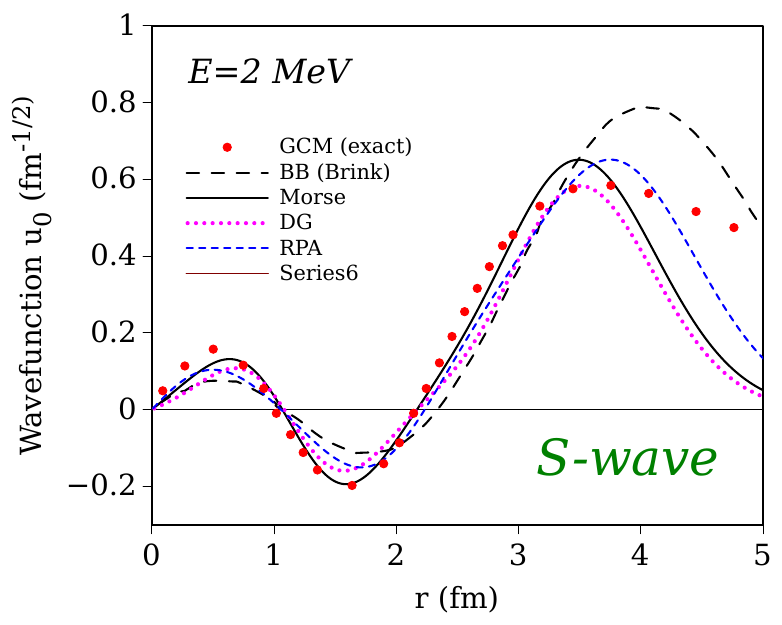}}
 \caption{Scattering wavefunction for \textit{S}-wave at 2 MeV in comparison with results of Yoshiharu \textit{et al.} \cite{yoshi}}
\label{S_r}
\end{figure}

\begin{figure}[H]
\centering
{\includegraphics[scale=0.8,angle=0]{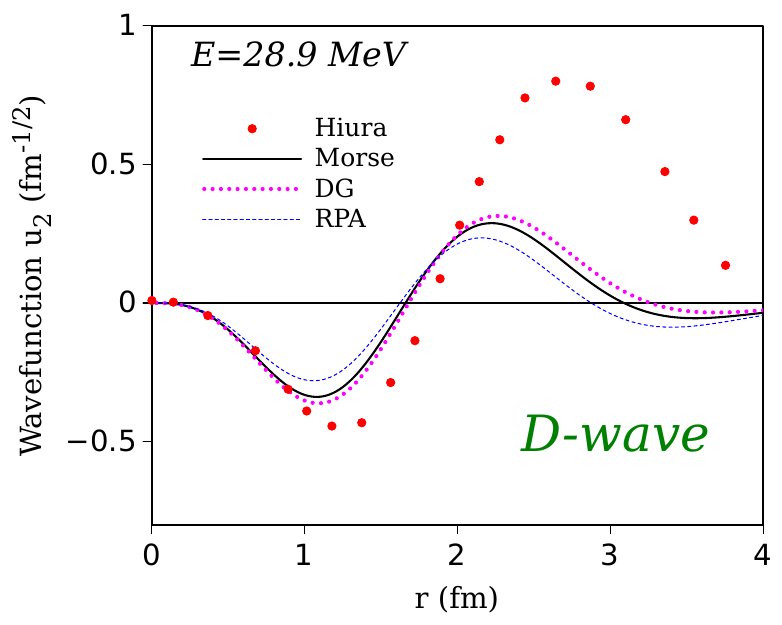}}
 \caption{Scattering wavefunction for \textit{S}-wave at 28.9 MeV in comparison with results of Yoshiharu \textit{et al.} \cite{yoshi}}
\label{S_r}
\end{figure}

\begin{figure}[H]
\centering
{\includegraphics[scale=0.8,angle=0]{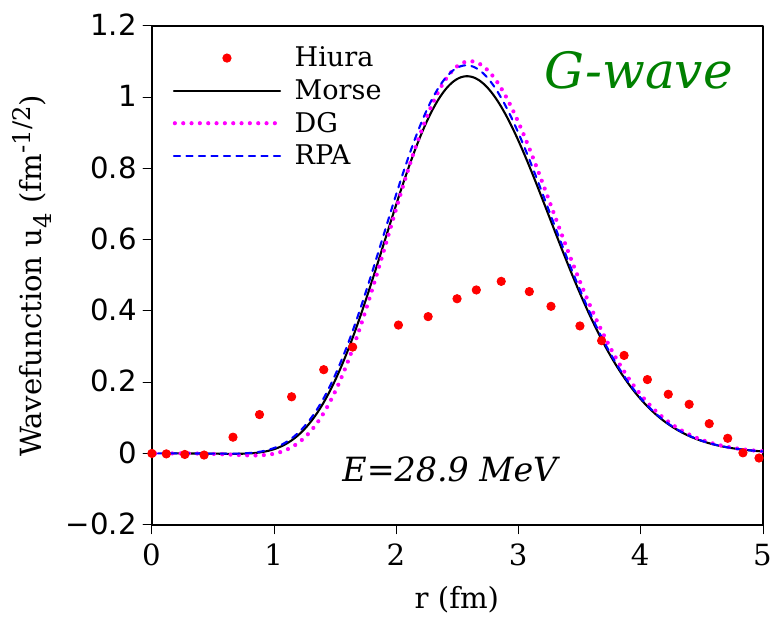}} \caption{Scattering wavefunction for \textit{S}-wave at 28.9 MeV in comparison with results of Yoshiharu \textit{et al.} \cite{yoshi}}
\label{S_r}
\end{figure}

\noindent For the s-wave ($\ell = 0$), the wavefunction displays the expected monotonic behavior
at short distances and a well-defined oscillatory structure at larger separations,
reflecting the dominance of the nuclear interaction at low angular momentum. In the
case of the d-wave ($\ell = 2$), the presence of the centrifugal barrier introduces
additional radial structure, leading to suppressed amplitudes at small radii and
enhanced oscillatory behavior beyond the interaction region. The g-wave ($\ell = 4$)
exhibits an even stronger centrifugal effect, resulting in comparatively smaller
amplitudes and a delayed onset of asymptotic oscillations. These trends are consistent
with the expected angular-momentum dependence of cluster--cluster scattering and
validate the numerical stability of the method across higher partial waves.

\noindent An important outcome of the present analysis is that the wavefunctions obtained using
the phase function formalism retain physical transparency while remaining
computationally efficient. Since the phase and amplitude functions are evolved
simultaneously, the method provides direct access to the internal structure of the
scattering solution, offering insight into how the interaction potential shapes the
radial behavior of the wavefunction. The use of a Morse-type nuclear interaction
combined with a finite-size Coulomb term ensures that both short-range and long-range
effects are treated in a unified manner, leading to well-behaved solutions over the
entire radial domain.

\noindent The phenomenological potentials we used has a repulsive core at
around R = 2 fm for the orbital angular momentum $\ell$= 0 and 2 waves and has been previously confirmed to originate from
the Pauli principle \cite{kanada} and similar repulsive core is observed in our potentials. Overall, the results confirm that the phase function method is not limited to the
determination of scattering phase shifts alone but can be reliably extended to construct
scattering wavefunctions for the $\alpha\alpha$ system. The approach offers a compact
and numerically stable framework for studying cluster scattering, making it
particularly suitable for applications where computational efficiency and direct
control over scattering observables are essential.

\subsection{Coulomb-modified effective range expansion}

The low-energy scattering parameters are extracted using the Coulomb-modified effective range expansion (ERE) \cite{el}

\begin{equation}
K_{\ell}(p)
=
C_{\eta,\ell}^{2} p^{2\ell+1}\cot\delta_{\ell}(p)
+
\gamma h_{\ell}(p)
=
-\frac{1}{a_{\ell}}
+
\frac{1}{2} r_{\ell} p^{2}
-
\frac{1}{4} P_{\ell} p^{4}
+
\mathcal{O}(p^{6}),
\label{eq:C1}
\end{equation}

where $a_\ell$, $r_\ell$, and $P_\ell$ denote the scattering length, effective range, and shape parameter, respectively. The Coulomb penetration factor is $
C_{\eta,\ell}^{2}
=
\frac{2^{2\ell}}{[(2\ell+1)!]^{2}}
\, C_{\eta,0}^{2}
\prod_{s=1}^{\ell}
(s^{2}+\eta^{2}),
$
with
$
C_{\eta,0}^{2}
=
\frac{2\pi\eta}{e^{2\pi\eta}-1},
$
and $\eta=\gamma/(2p)$, where
\(
\gamma = 2\mu\alpha_{\rm EM} Z_1 Z_2
\).
The function $h_\ell(p)$ is given by
$
h_{\ell}(p)
=
p^{2\ell}
\frac{C_{\eta,\ell}^{2}}{C_{\eta,0}^{2}}
\left[
\mathrm{Re}\,\psi(i\eta)
-
\ln |\eta|
\right].
$ 
\begin{table}[H]
\label{scatparam}
\centering
\caption{Comparison of scattering parameters $a_0$, $r_0$ \& $P_0$ obtained using different interaction models for S-wave. Our results are in close agreement with those in literature \cite{rasche}\cite{el}. Binding energy was calculated for DG \textit{S}-wave by using Marsiglo's matrix method \cite{marsiglo}.}

\renewcommand{\arraystretch}{0.9}

\begin{tabular}{lccc|ccc}
\hline
& \multicolumn{3}{c|}{\textbf{OUR}} & \multicolumn{3}{c}{\textbf{Literature}} \\

\cline{2-7}

\textbf{Parameter} & \textbf{Morse} & \textbf{DG} & \textbf{RPA} & \textbf{NLO} & \textbf{NNLO} & \textbf{Empirical} \cite{rasche} \\
\hline
$a_0$ & $-919.123$ & $-1831.181$ & $-1800.747$ & $-1800$ & $-1500$ & $-1650$ \\
$r_0$ & $1.165$ & $1.083$ & $1.075$ & $1.045$ & $1.061$ & $1.08$ \\
$P_0$ & $-1.15$ & $-1.837$ & $-1.934$ & $-2.297$ & $-2.277$ & $-1.76$\\ \hline
$\text{Binding~Energy}$ & $-0.138$ & $ 0.103$ & $0.073$ & $-$ & $-$ & $0.0918$~\cite{exp} \\

\hline
\end{tabular}
\label{scatparam}
\end{table}

S-wave ERE parameters $a_0$, $r_0$ and $P_0$, collected in table~\ref{scatparam}. The fit range to determine these is from $E_{\mathrm{Lab}} = 0.01$ to $5.0\,\mathrm{MeV}$ for all interactions. We see that these parameters are consistent with the empirical determinations, but they are also afflicted with sizeable uncertainties. Note that there is sensitivity to the fit range as well as to the position of the $0^{+}$ resonance, the $^{8}\mathrm{Be}$ ground state, as discussed in ref.~\cite{Buck}. In our calculation, $^{8}\mathrm{Be}$ is very weakly bound. This appears to be in contradiction to the scattering lengths given in table~3, but these values are very sensitive to the fitting range employed to extract them, see also ref.~\cite{el}.
\section{Conclusion}
\noindent In the present work, the phase function method has been successfully extended beyond
its conventional application to scattering phase shifts in order to construct nonantisymmetrized wavefunctions for the $\alpha\alpha$ system. By employing genetically optimized
interaction parameters obtained from earlier phase-shift analyses, wavefunctions for
the $\ell = 0$, 2, and 4 partial waves have been computed within a unified and
numerically stable framework. The formulation avoids the direct solution of the
second-order Schr\"odinger equation and instead relies on first-order differential
equations for the phase and amplitude functions, thereby reducing computational
complexity while preserving the essential physical content of the scattering problem.

\noindent The resulting wavefunctions exhibit physically consistent behavior over the entire
radial domain, including smooth evolution in the interaction region and correct
asymptotic oscillatory behavior at large distances. The angular-momentum dependence of
the solutions is clearly manifested through the increasing influence of the
centrifugal barrier from s- to g-waves, demonstrating that the method reliably captures
the characteristic features of cluster-cluster scattering dynamics. The stabilization
of the amplitude function at large radii further confirms the internal consistency of
the approach and the suitability of the adopted interaction model.

\noindent A significant outcome of the present analysis is that the computed wavefunctions show
very good agreement with previously reported results by Hiura \textit{et al.}, thereby
providing an important validation of the present phase function based formulation.
This agreement indicates that the essential physics of the $\alpha\alpha$ interaction
is effectively encoded in the optimized potential and the phase evolution, even
without explicitly invoking more elaborate microscopic formalisms. The consistency
with established results strengthens confidence in the accuracy and reliability of
the present computational approach.

\noindent The use of a Morse-type nuclear interaction in conjunction with a finite-size Coulomb
potential ensures a realistic and well-behaved description of both short-range nuclear
attraction and long-range electrostatic repulsion. This combination contributes to the
numerical stability of the solutions and allows for a transparent interpretation of
the radial structure of the scattering wavefunctions. The ability to reconstruct
wavefunctions directly from phase evolution highlights the strength of the phase
function method as a unified tool for scattering studies.

\noindent Overall, this investigation demonstrates that the phase function framework provides an
efficient and physically meaningful approach for analyzing $\alpha\alpha$ scattering,
yielding scattering amplitudes and wavefunctions in addition to phase shifts within a
single formalism. The methodology developed here can be readily extended to other
cluster and nucleus--nucleus scattering systems and to different interaction models,
making it a promising tool for future studies of nuclear scattering and inverse
problems.
    
\section*{Data and Code Availibility} The numerical data and computational codes used in the study (including simulation and analysis routines) are available from the corresponding author upon reasonable request. 

\section*{Acknowledgment}
The author acknowledges the institutional support and research incentives provided by Chandigarh Engineering College, CGC University, Mohali.

\vspace{0.2cm}


\let\doi\relax

\appendix

\section{Appendix: Explicit Forms of the Amplitude and wavefunctions}
\noindent For completeness, we provide below the fully expanded expressions used in the numerical implementation for the $\ell = 0, 2,$ and $4$ partial waves. The equations follow directly from the general phase-amplitude formalism presented in Sec.~\ref{Sec2}.
\subsection{Amplitude Equations}

\noindent The general amplitude equation is
\begin{equation}
A_\ell'(r)
=
-\frac{A_\ell(r)\,V(r)}{k\left(\hbar^2/2\mu\right)}
\Big[
\cos\delta_\ell(r)\,\hat{j}_\ell(kr)
-
\sin\delta_\ell(r)\,\hat{\eta}_\ell(kr)
\Big]
\Big[
\sin\delta_\ell(r)\,\hat{j}_\ell(kr)
+
\cos\delta_\ell(r)\,\hat{\eta}_\ell(kr)
\Big].
\end{equation}

Below we list the explicit forms.

\paragraph*{$\ell = 0$}

Using
\[
\hat{j}_0(kr)=\sin(kr), 
\qquad
\hat{\eta}_0(kr)=-\cos(kr),
\]
we obtain
\begin{align}
A_0'(r)
&=
-\frac{A_0 V(r)}{k\left(\hbar^2/2\mu\right)}
\left[
\cos\delta_0 \sin(kr)
+
\sin\delta_0 \cos(kr)
\right]
\nonumber\\
&\quad\times
\left[
\sin\delta_0 \sin(kr)
-
\cos\delta_0 \cos(kr)
\right].
\label{A0_appendix}
\end{align}

\paragraph*{$\ell = 2$}

For
\begin{align*}
\hat{j}_2(kr)
&=
\left(\frac{3}{(kr)^2}-1\right)\sin(kr)
-
\frac{3}{kr}\cos(kr), \\
\hat{\eta}_2(kr)
&=
-\left(\frac{3}{(kr)^2}-1\right)\cos(kr)
-
\frac{3}{kr}\sin(kr),
\end{align*}
the amplitude equation becomes
\begin{align}
A_2'(r)
&=
-\frac{A_2 V(r)}{k\left(\hbar^2/2\mu\right)}
\Big[
\cos\delta_2\,\hat{j}_2(kr)
-
\sin\delta_2\,\hat{\eta}_2(kr)
\Big]
\nonumber\\
&\quad\times
\Big[
\sin\delta_2\,\hat{j}_2(kr)
+
\cos\delta_2\,\hat{\eta}_2(kr)
\Big].
\label{A2_appendix}
\end{align}

\paragraph*{$\ell = 4$}

For
\begin{align*}
\hat{j}_4(kr)
&=
\left(
\frac{105}{(kr)^4}
-
\frac{45}{(kr)^2}
+
1
\right)\sin(kr)
+
\left(
\frac{10}{kr}
-
\frac{105}{(kr)^3}
\right)\cos(kr), \\
\hat{\eta}_4(kr)
&=
-\left(
\frac{105}{(kr)^4}
-
\frac{45}{(kr)^2}
+
1
\right)\cos(kr)
+
\left(
\frac{10}{kr}
-
\frac{105}{(kr)^3}
\right)\sin(kr),
\end{align*}
we obtain
\begin{align}
A_4'(r)
&=
-\frac{A_4 V(r)}{k\left(\hbar^2/2\mu\right)}
\Big[
\cos\delta_4\,\hat{j}_4(kr)
-
\sin\delta_4\,\hat{\eta}_4(kr)
\Big]
\nonumber\\
&\quad\times
\Big[
\sin\delta_4\,\hat{j}_4(kr)
+
\cos\delta_4\,\hat{\eta}_4(kr)
\Big].
\label{A4_appendix}
\end{align}

\subsection{wavefunction Expressions}\label{wavefunctions_all}

The reconstructed reduced radial wavefunction is

\begin{equation}
u_\ell(r)
=
A_\ell(r)
\Big[
\cos\delta_\ell(r)\,\hat{j}_\ell(kr)
-
\sin\delta_\ell(r)\,\hat{\eta}_\ell(kr)
\Big].
\end{equation}

Explicitly:

\paragraph*{$\ell = 0$}

\begin{equation}
u_0(r)
=
A_0(r)
\left[
\cos\delta_0(r)\sin(kr)
+
\sin\delta_0(r)\cos(kr)
\right].
\label{u0_appendix}
\end{equation}

\paragraph*{$\ell = 2$}

\begin{equation}
u_2(r)
=
A_2(r)
\Big[
\cos\delta_2(r)\,\hat{j}_2(kr)
-
\sin\delta_2(r)\,\hat{\eta}_2(kr)
\Big].
\label{u2_appendix}
\end{equation}

\paragraph*{$\ell = 4$}

\begin{equation}
u_4(r)
=
A_4(r)
\Big[
\cos\delta_4(r)\,\hat{j}_4(kr)
-
\sin\delta_4(r)\,\hat{\eta}_4(kr)
\Big].
\label{u4_appendix}
\end{equation}
\section{Matrix Mechanics Formulation}
In order to extract the bound and quasi-bound state energies corresponding to the $\alpha$--$\alpha$ interaction potential obtained via the Phase Function Method (PFM), we employ a numerical matrix mechanics approach. This method transforms the radial Schrodinger equation into a matrix eigenvalue problem, allowing efficient computation of energy eigenvalues and corresponding wavefunctions for arbitrary potentials.
\subsection{Radial Schrödinger Equation}
For a central potential, the total wavefunction can be separated as
\begin{equation}
\psi(\mathbf{r}) = R(r) Y_{\ell}^{m}(\theta,\phi),
\end{equation}
where $Y_{\ell}^{m}(\theta,\phi)$ are the spherical harmonics. Defining the reduced radial wavefunction
\begin{equation}
u(r) = r R(r),
\end{equation}
the Schrödinger equation reduces to a one-dimensional radial form:
\begin{equation}
-\frac{\hbar^2}{2\mu} \frac{d^2 u(r)}{dr^2} + V_{\text{eff}}(r)\,u(r) = E\,u(r),
\end{equation}
where $\mu$ is the reduced mass of the $\alpha$--$\alpha$ system and the effective potential (Double Gaussian (DG) potential) is given by
\begin{equation}
V_{\text{eff}}(r) = V_r.e^{-\mu_r^2 r^2}-V_a.e^{-\mu_a^2 r^2}+V_0\frac{e^{-r/a}}{(1-e^{-r/a})}
\end{equation}
\subsection{Operator Formulation}

The radial Schrödinger equation can be expressed in operator form as
\begin{equation}
\left[H_0 + V_{\text{eff}}(r)\right] |u\rangle = E |u\rangle,
\end{equation}
where $H_0$ represents the kinetic energy operator together with an embedding infinite square well potential, introduced to define a convenient basis.

\subsection{Choice of Basis}

To construct the matrix representation, the system is confined within an infinite spherical well of radius $a$, such that
\begin{equation}
u(0) = 0, \quad u(a) = 0.
\end{equation}
The corresponding orthonormal basis functions are
\begin{equation}
\phi_n(r) = \sqrt{\frac{2}{a}} \sin\left(\frac{n\pi r}{a}\right),
\end{equation}
which satisfy
\begin{equation}
H_0 |\phi_n\rangle = E_n^{(0)} |\phi_n\rangle,
\end{equation}
with eigenvalues
\begin{equation}
E_n^{(0)} = \frac{n^2 \pi^2 \hbar^2}{2\mu a^2}.
\end{equation}

\subsection{Expansion of the Wavefunction}

The unknown radial wavefunction is expanded in terms of this complete basis:
\begin{equation}
|u\rangle = \sum_{m=1}^{\infty} c_m |\phi_m\rangle,
\end{equation}
where $c_m$ are expansion coefficients to be determined.

\subsection{Matrix Eigenvalue Equation}

Substituting the above expansion into the operator equation and projecting onto the basis state $\langle \phi_n |$, we obtain
\begin{equation}
\sum_{m=1}^{\infty} H_{nm} c_m = E c_n,
\end{equation}
which represents a matrix eigenvalue problem. The Hamiltonian matrix elements are defined as
\begin{equation}
H_{nm} = \langle \phi_n | H | \phi_m \rangle.
\end{equation}

Explicitly, these matrix elements can be written as
\begin{equation}
H_{nm} = \delta_{nm} E_n^{(0)} + \frac{2}{a} \int_0^a
\sin\left(\frac{n\pi r}{a}\right)
\left[
V_{\text{PFM}}(r) + \frac{\hbar^2 \ell(\ell+1)}{2\mu r^2}
\right]
\sin\left(\frac{m\pi r}{a}\right)\, dr,
\end{equation}
where $\delta_{nm}$ is the Kronecker delta.

\subsection{Numerical Implementation}

In practical calculations, the infinite matrix is truncated to a finite size $n_{\text{max}}$, yielding a matrix equation of the form
\begin{equation}
\mathbf{H} \mathbf{c} = E \mathbf{c}.
\end{equation}
The eigenvalues $E$ correspond to the energy levels of the system, while the eigenvectors $\mathbf{c}$ determine the radial wavefunctions via
\begin{equation}
u(r) = \sum_{m=1}^{n_{\text{max}}} c_m \phi_m(r).
\end{equation}

\subsection{Physical Interpretation and Applicability}

This formulation converts the differential Schrödinger equation into a linear algebra problem, which can be efficiently solved using standard numerical diagonalization techniques. The method is particularly advantageous for potentials such as $V_{\text{PFM}}(r)$, which may not admit analytical solutions.

The accuracy of the results depends on the choice of the cutoff radius $a$ and the matrix size $n_{\text{max}}$. The parameter $a$ must be sufficiently large to encompass the spatial extent of the wavefunction, while $n_{\text{max}}$ controls the resolution of the basis. Convergence is achieved by systematically increasing both parameters.

This approach enables the determination of bound and quasi-bound states of the $\alpha$--$\alpha$ system and provides a consistent framework to analyze the energy spectrum corresponding to the interaction potential obtained via the Phase Function Method.

\end{document}